\documentclass[a4paper,12pt]{article}
\usepackage[english]{babel}
\usepackage{jheppub2}
\usepackage{subfig}
\usepackage[T1]{fontenc} 
\usepackage{slashed}
\usepackage{hyperref}
\usepackage{flushend}
\usepackage{graphicx}
\usepackage{epsfig}
\usepackage{amsmath}
\usepackage{amssymb}
\usepackage{float}
\usepackage{placeins}
\usepackage{braket}

\allowdisplaybreaks[4]

\title{OPE and a low-energy theorem in QCD-like theories}

\author[a,b]{Matteo Becchetti}
\author[b]{Marco Bochicchio}

\affiliation[a]{Sapienza - Universit\`a di Roma, Dipartimento di Fisica, Piazzale Aldo Moro 5, 00185, Rome, Italy}
\affiliation[b]{INFN Sezione di Roma, Piazzale Aldo Moro 2, 00185, Rome, Italy}

\emailAdd{matteo.becchetti@roma1.infn.it}
\emailAdd{marco.bochicchio@roma1.infn.it}

\abstract{We verify, both perturbatively and nonperturbatively asymptotically in the ultraviolet (UV), a special case of a low-energy theorem of the NSVZ type in QCD-like theories, recently derived in \href{https://doi.org/10.1103/PhysRevD.95.054010}{Phys. Rev. D {\bf 95} (2017) 054010}, that relates the logarithmic derivative with respect to the gauge coupling, or the logarithmic derivative with respect to the renormalization-group (RG) invariant scale,
of an $n$-point correlator of local operators in one side to an $n+1$-point correlator with the insertion of $\Tr F^2$ at zero momentum in the other side. 
Our computation involves the operator product expansion (OPE) of the scalar glueball operator, $\Tr F^2$, in massless QCD, worked out perturbatively in \href{https://doi.org/10.1007/JHEP12(2012)119}{JHEP {\bf 1212} (2012) 119} -- and in its RG-improved form in the present paper --
by means of which we extract both the perturbative divergences and the nonperturbative UV asymptotics in both sides. 
We also discuss the role of the contact terms in the OPE, both finite and divergent, discovered some years ago in \href{https://doi.org/10.1007/JHEP12(2012)119}{JHEP {\bf 1212} (2012) 119},
in relation to the low-energy theorem. Besides, working the other way around by assuming the low-energy theorem for any 2-point correlator of a multiplicatively renormalizable gauge-invariant operator, we compute in a massless QCD-like theory the corresponding perturbative OPE to the order of $g^2$ and nonperturbative asymptotics.
The low-energy theorem has a number of applications: to the renormalization in asymptotically free QCD-like theories, both perturbatively and nonperturbatively in the large-$N$ 't Hooft and Veneziano expansions, and to the way the open/closed string duality may or may not be realized in the would-be solution by canonical string theories for QCD-like theories, both perturbatively and in the 't Hooft large-$N$ expansion. Our computations will also enter further developments based on the low-energy theorem.}

\DeclareMathOperator{\Tr}{Tr}
\newcommand{\be}{\begin{equation}}
\newcommand{\ee}{\end{equation}}
\newcommand{\nn}{\nonumber}
\newcommand{\bea}{\begin{eqnarray}}
\newcommand{\eea}{\end{eqnarray}}
\newcommand{\bfig}{\begin{figure}}
\newcommand{\efig}{\end{figure}}
\newcommand{\bc}{\begin{center}}
\newcommand{\ec}{\end{center}}

\newcommand{\f}[2]{\frac{#1}{#2}}
 
\newcommand{\cf}{C_{\scriptscriptstyle{F}}} 
\newcommand{\ca}{C_{\scriptscriptstyle{A}}}
\newcommand{\tr}{T_{\scriptscriptstyle{F}}}

\newcommand{\Nf}{N_{\scriptscriptstyle{f}}}

\newcommand{\as}{a_{\scriptscriptstyle{s}}}

\begin{document}

\maketitle
\flushbottom

\section{Introduction}

\subsection{Physics motivations}

One of the aims of the present paper is to verify, both perturbatively and nonperturbatively asymptotically in the ultraviolet (UV), a special case of a recently derived low-energy theorem \cite{MBR} of the Novikov-Shifman-Vainshtein-Zakharov (NSVZ) type \cite{NSVZ} in SU(N) QCD-like gauge theories. \par
It relates the logarithmic derivative with respect to the 't Hooft gauge coupling, $g^2=g^2_{YM} N$, of an $n$-point correlator of local operators, $\mathcal{O}_k$, to an $n+1$-point correlator with the insertion of $\Tr \mathcal{F}^2 $ at zero momentum \cite{MBR}:
\bea
\label{LET}
\f{\partial\langle \mathcal{O}_1 \cdots \mathcal{O}_n \rangle}{\partial \log g} = \frac{N}{g^2} \int \langle \mathcal{O}_1\cdots \mathcal{O}_n \Tr \mathcal{F}^2(x)\rangle- \langle\mathcal{O}_1\cdots \mathcal{O}_n\rangle \langle \Tr \mathcal{F}^2(x)\rangle  d^4x
\eea
where the Wilsonian normalization of the Yang-Mills (YM) action is chosen (Subsec. \ref{2.1}) and the operators, $\Tr \mathcal{F}^2 $ and $\mathcal{O}_k$, are $g$ independent (Subsec. \ref{2.1}).\par
It admits another version \cite{MBR}, where the logarithmic derivative with respect to the gauge coupling is replaced by the logarithmic derivative with respect to the renormalization-group (RG) invariant scale, $\Lambda_{QCD}$, in an asymptotically free (AF) QCD-like theory (Subsec. \ref{2.2}):
\bea
\label{LET0}
\dfrac{\partial \langle \mathcal{O}_1\dots \mathcal{O}_n \rangle}{\partial \log \Lambda_{QCD}} = -\frac{N\beta(g)}{g^3}\int \langle \mathcal{O}_1 \dots \mathcal{O}_n \Tr \mathcal{F}^2(x) \rangle - \langle \mathcal{O}_1 \dots \mathcal{O}_n \rangle \langle \Tr \mathcal{F}^2(x) \rangle  d^4x
\eea
After rescaling the gauge fields in the functional integral by a factor of $\frac{g}{\sqrt N}$, the low-energy theorem admits a trivially equivalent canonical version with the canonical normalization of the YM action (Subsec. \ref{2.3}):
\bea
\label{LowEneT}
&&\big(\sum^{k=n}_{k=1} c_k \big)  \langle O_1\dots O_n \rangle +  \dfrac{\partial \langle O_1\dots O_n \rangle}{\partial \log g} \nonumber \\
&&= \int \langle O_1 \dots O_n \Tr F^2(x) \rangle - \langle O_1 \dots O_n \rangle \langle \Tr F^2(x) \rangle d^4x
\eea
in terms of operators, $ \Tr F^2$ and $O_k$, defined by the very same rescaling, satisfying $\frac{g^2}{N} \Tr F^2= \Tr \mathcal{F}^2$ and $(\frac{g}{\sqrt N})^{c_k} O_k= \mathcal{O}_k$ for some $c_k$. The canonically normalized operators, $\Tr F^2$ and $O_k$, depend now on $g$ (Subsec. \ref{2.3}). Eq. \eqref{LET0} also admits a canonical version (Subsec. \ref{2.3}): 
\bea
\label{LowEneT1}
&&- \frac{\beta(g)}{g}  \big(\sum^{k=n}_{k=1} c_k \big)  \langle O_1\dots O_n \rangle + \dfrac{\partial \langle O_1\dots O_n \rangle}{\partial \log \Lambda_{QCD}} \nonumber \\
&&= - \frac{\beta(g)}{g} \int \langle O_1 \dots O_n \Tr F^2(x) \rangle - \langle O_1 \dots O_n \rangle \langle \Tr F^2(x) \rangle d^4x
\eea
The canonical version of Eq. \eqref{LET}, which is most suitable for perturbative computations, -- Eq. \eqref{LowEneT} for $n=2$ and $O_k= \Tr F^2$ -- has been employed to analyze the renormalization properties of QCD-like theories perturbatively to the order of $g^2$ in \cite{MBL}
and the way the open/closed string duality \cite{V,DV} may be actually implemented \cite{MBL} in string theories realizing perturbatively \cite{DV} QCD-like theories.\par
The second version -- Eq. \eqref{LET0} with $\mathcal{O}_k= \Tr \mathcal{F}^2$ --  has been employed to compute the nonperturbative countertems \cite{MBR} in the large-$N$ 't Hooft \cite{H1} and Veneziano \cite{V0} expansions of QCD-like theories. \par 
Moreover, it has entered crucially a no-go theorem \cite{MBL} that the nonperturbative renormalization in the 't Hooft large-$N$ QCD S matrix is incompatible with the open/closed string duality of a would-be canonical string solution, which therefore does not exist \cite{MBL}. \par
By a canonical string solution we mean \cite{MBL} a perturbative expansion in the string coupling, $g_s \sim \frac{1}{N}$, for the string S matrix that matches the topology of the 't Hooft large-$N$ expansion and computes the large-$N$ QCD S matrix by means of an auxiliary 2d conformal field theory living on the string world-sheet with fixed topology \cite{H1,V0,VR,Mal}. \par
A noncanonical way-out to the no-go theorem has been suggested in \cite{MBL,MBG}. Relatedly, the large-$N$ Veneziano expansion \cite{V0} has been discussed in \cite{MBR,MBL,MBG}. \par
Because of the importance of the low-energy theorem for the above subjects, a deeper understanding and an explicit evaluation of both sides in Eqs. \eqref{LowEneT} and \eqref{LowEneT1} are most interesting. \par

\subsection{Plan of the paper} 

In Sec. \ref{2} we recall the proof of various versions of the low-energy theorem in \cite{MBR}.\par
In Sec. \ref{20} we describe the rationale behind our computations based on the operator product expansion (OPE) in relation to the low-energy theorem.\par
In Subsec. \ref{32} we summarize the result of the computation in \cite{Ch,Z1,Z2}, which we employ in the present paper, of the perturbative OPE coefficients for $\Tr F^2(x) \Tr F^2(0)$
(Eqs. \eqref{OPEexp}, \eqref{C0xstr} and \eqref{C1xstr}) in QCD with massless quarks (massless QCD for short). \par
In Sec. \ref{3} we verify the low-energy theorem perturbatively to the order of $g^4$ for $n=2$ and $O_k= \Tr F^2$ on the basis of the aforementioned OPE in \cite{Ch,Z1,Z2}.\par
In fact, the divergent parts in both sides of Eq. \eqref{LowEneT} for $n=2$ and $O_k= \Tr F^2$ have already been computed to the order of $g^2$ in perturbation theory in \cite{MBL}, thus partially verifying a special case of the low-energy theorem perturbatively. \par
Yet, we include (Subsec. \ref{3.1}) in the aforementioned computation the finite contact term to the order of $g^0$ in $C_1^{(S)}$ (Eq. \eqref{C1xstr}) \cite{Ch}, which arises from performing the OPE in Eq. \eqref{LowEneT}, that has been skipped in \cite{MBL}. \par
Moreover, we extend the perturbative computation in \cite{MBL} to the order of $g^4$ in Subsec. \ref{3.2}, including as well the divergent contact term to the order of $g^4$ in $C_1^{(S)}$ (Eq. \eqref{C1xstr}) discovered some years ago in \cite{Ch}, whose renormalization has been recently discussed in \cite{Z2}. \par
In Sec. \ref{4} we compute the nonperturbative universal, i.e., renormalization-scheme independent, UV asymptotics \footnote{We define in App. \ref{A} what we mean by the universal UV asymptotics.} in the coordinate representation in both sides of Eq. \eqref{LowEneT1} for $n=2$ and $O_k= \Tr F^2$.\par
Our nonperturbative computation furnishes a detailed derivation and an improvement to the next to leading logs of the crucial -- for the no-go theorem 
in \cite{MBL} -- nonperturbative UV leading-log asymptotic estimate in the coordinate representation in \cite{MBL}, and provides another check of the low-energy theorem. \par
In order to perform the computation, we employ firstly the perturbative OPE of $\Tr F^2(x) \Tr F^2(0)$ worked out in \cite{Ch,Z1,Z2} that is recalled in App. \ref{B}. 
A previous perturbative lower-order computation appeared in \cite{K}. 
We pass from the perturbative OPE in the momentum representation \cite{Ch,Z1,Z2} in Apps. \ref{B1} and \ref{B4} to the OPE in the coordinate representation by the Fourier transform in Apps. \ref{B2} and \ref{B5}. 
From it we get the normalization for the nonperturbative universal RG-improved UV asymptotics of the OPE coefficients in the coordinate representation derived a priori in App. \ref{A}, which includes the leading and next to leading logs. \par
Moreover, by making the OPE coefficients RG invariant by a suitable rescaling of the operators, we verify in Apps. \ref{B3} and \ref{B6} their nonperturbative UV asymptotics by rewriting the perturbative computation, originally expressed in terms of $g(\mu) $ and large logs \cite{Ch,Z1,Z2}, in terms of the running coupling, $g(x)$, to the given perturbative order.\par
In passing, for future applications, we compute in App. \ref{B3} the nonperturbative UV asymptotics in massless QCD of the 2-point correlator of $\frac{g^2}{N} \Tr F^2$ -- the YM Lagrangian density with the canonical normalization (Subsec. \ref{2.3}) -- which coincides with the continuum limit of the 2-point correlator of the Wilson plaquette on the lattice (Subsec. \ref{2.3}).
As the operator $\frac{g^2}{N}  \Tr F^2$ is not RG invariant, our computation also includes the scale-dependent corrections to the universal UV asymptotics (App. \ref{B3}). \par
Previous related results about the universal asymptotics of the OPE both in the momentum and coordinate representation have been obtained in \cite{MBM,MBN} and, only about the leading logs, in the momentum representation in \cite{R} and in the coordinate representation in \cite{MBL}. \par
In Sec. \ref{5}, working the other way around by assuming the low-energy theorem for any 2-point correlator, $\langle O(z)  O(0) \rangle $, of a canonically normalized multiplicatively renormalizable gauge-invariant operator $O$, we compute the corresponding perturbative OPE to the order of $g^2$ and nonperturbative UV asymptotics. \par 
In Sec. \ref{7} we summarize our conclusions. Our computations will also enter further developments that assume the low-energy theorem. \par

 \section{The low-energy theorem} \label{2}

\subsection{Low-energy theorem in terms of the Wilsonian coupling} \label{2.1}

For completeness we report the proof of the low-energy theorem in QCD-like theories according to \cite{MBR}.\par
Given a set of local operators, $\mathcal{O}_k$, and the Wilsonian normalization of the YM action, by deriving:
\be \label{Alfa}
\langle \mathcal{O}_1\cdots\mathcal{O}_n\rangle = \f{\int \mathcal{O}_1\cdots\mathcal{O}_n e^{-\f{N}{2 g^2}\int \Tr \mathcal{F}^2(x)d^4x + \cdots}}{e^{-\f{N}{2 g^2}\int \Tr \mathcal{F}^2(x)d^4x + \cdots}}
\ee
with respect to $-\f{1}{g^2}$, we obtain:
\be
\label{Intro1}
\f{\partial\langle \mathcal{O}_1 \cdots \mathcal{O}_n\rangle}{\partial \log g} = \f{N}{g^2} \int \langle \mathcal{O}_1\cdots \mathcal{O}_n \Tr \mathcal{F}^2(x)\rangle- \langle\mathcal{O}_1\cdots \mathcal{O}_n\rangle\langle \Tr \mathcal{F}^2(x)\rangle d^4x
\ee
where the trace, $\Tr$, is in the fundamental representation, $\Tr \mathcal{F}^2 \equiv \Tr (\mathcal{F}_{\mu\nu}\mathcal{F}^{\mu\nu})$, the sum over repeated indices is understood, and $\mathcal{F}_{\mu\nu} = \partial_{\mu}A_{\nu} - \partial_{\nu} A_{\mu} + i [A_{\mu},A_{\nu}]$. \par
From the derivation it is clear that all the operators -- $\mathcal{O}_k$, $\Tr \mathcal{F}^2$, and in the dots -- are chosen to be $g$ independent. It also is clear that $g$ in Eqs. \eqref{Alfa} and \eqref{Intro1} is the bare coupling. \par 
Interestingly, the Wilsonian normalization of the YM action in Eq. \eqref{Alfa} also occurs in nonperturbative computations in lattice gauge theories. \par

\subsection{Low-energy theorem in terms of $\Lambda_{QCD}$} \label{2.2}

A second version \cite{MBR} of the low-energy theorem holds in an AF QCD-like theory:
\be
\label{LETlam}
\dfrac{\partial \langle \mathcal{O}_1\dots \mathcal{O}_n \rangle}{\partial \log \Lambda_{QCD}} = -\frac{N\beta(g)}{g^3}\int \langle \mathcal{O}_1 \dots \mathcal{O}_n \Tr \mathcal{F}^2(x) \rangle - \langle \mathcal{O}_1 \dots \mathcal{O}_n \rangle \langle \Tr \mathcal{F}^2(x) \rangle  d^4x
\ee
as it follows by employing the chain rule, $\frac{\partial}{\partial \log g}=\frac{\partial \Lambda_{QCD}}{\partial \log g} \frac{ \partial}{\partial \Lambda_{QCD}} $, the defining relation:
\be
\left(\f{\partial}{\partial \log \Lambda} + \beta(g)\f{\partial}{\partial g}\right)\Lambda_{QCD} = 0
\ee
and the identity:
\be
\f{\partial \Lambda_{QCD}}{\partial \log \Lambda} = \Lambda_{QCD}
\ee
since $\Lambda_{QCD} = e^{\log \Lambda}f(g)$ for a function $f(g)$.

\subsection{Low-energy theorem in terms of the canonical coupling} \label{2.3}

In order to verify the low-energy theorem in perturbation theory it is convenient to employ the canonical normalization of the YM action \cite{MBL}.\par
Thus, we rescale the gauge fields in the functional integral by a factor of $\f{g}{\sqrt N}$. Of course, the rescaling does not affect the vev of the operators, as it is just a change of variables \footnote{Certainly, this statement holds to every order of perturbation theory. It also holds nonperturbatively provided that the theory is regularized, as for example in dimensional regularization, in such a way that no rescaling anomaly arises \cite{Arkani}. Moreover, it holds nonperturbatively as well if the theory is regularized on a finite lattice, since then the functional integral is finite dimensional and no rescaling anomaly may arise.}.\par Therefore, defining after the rescaling, $\frac{g^2}{N} \Tr F^2= \Tr \mathcal{F}^2$ and $(\frac{g}{\sqrt N})^{c_k} O_k= \mathcal{O}_k$ for some $c_k$, where now $F_{\mu\nu} = \partial_{\mu}A_{\nu} - \partial_{\nu} A_{\mu} + i \f{g}{\sqrt N}[A_{\mu},A_{\nu}]$ and $O_k$ are $g$ dependent but canonically normalized, we obtain the identity:
\bea \label{res}
\langle \mathcal{O}_1 \cdots \mathcal{O}_n\rangle = \prod^{k=n}_{k=1}(\frac{g}{\sqrt N})^{c_k}   \langle O_1\dots O_n \rangle
\eea
where the vev in the lhs is defined with the Wilsonian normalization in Eq. \eqref{Alfa} and in the rhs with the canonical normalization, i.e., after the aforementioned rescaling of the gauge fields in the functional integral in Eq. \eqref{Alfa}.\par
Interestingly, for $\Tr \mathcal{F}^2$ -- the YM Lagrangian density with the Wilsonian normalization, which coincides with the continuum limit of the Wilson plaquette on the lattice -- Eq. \eqref{res} reduces to:
\bea \label{res1}
\langle  \Tr \mathcal{F}^2 \cdots  \Tr \mathcal{F}^2 \rangle =  \langle  \frac{g^2}{N} \Tr F^2 \cdots  \frac{g^2}{N} \Tr F^2  \rangle 
\eea
since the Wilsonian normalization of the YM action occurs in lattice gauge theories as well (Subec. \ref{2.1}). \par
As a consequence, after the rescaling, Eq. \eqref{Intro1} reads:
\bea
\label{LowEneT0}
&&\big(\sum^{k=n}_{k=1} c_k \big)  \langle O_1\dots O_n \rangle +  \dfrac{\partial \langle O_1\dots O_n \rangle}{\partial \log g} \nonumber \\
&&= \int \langle O_1 \dots O_n \Tr F^2(x) \rangle - \langle O_1 \dots O_n \rangle \langle \Tr F^2(x) \rangle d^4x
\eea
Similarly, Eq. \eqref{LETlam} becomes:
\bea
\label{LowEneT10}
&&- \frac{\beta(g)}{g}  \big(\sum^{k=n}_{k=1} c_k \big)  \langle O_1\dots O_n \rangle + \dfrac{\partial \langle O_1\dots O_n \rangle}{\partial \log \Lambda_{QCD}} \nonumber \\
&&= - \frac{\beta(g)}{g} \int \langle O_1 \dots O_n \Tr F^2(x) \rangle - \langle O_1 \dots O_n \rangle \langle \Tr F^2(x) \rangle d^4x
\eea

\section{Main technical arguments} \label{20}

\subsection{Verifying Eq. \eqref{LowEneT} by the perturbative OPE} \label{000}

Eqs. \eqref{LowEneT} and \eqref{LowEneT0} apply to any product, $O_1 \dots O_n$, of bare -- canonically normalized -- local operators, not necessarily gauge invariant.\par
In order to verify perturbatively Eq. \eqref{LowEneT}, we assume that such local operators have a well defined -- in general nonvanishing -- anomalous dimension (Sec. \ref{5}). \par
Moreover, for the explicit computations in the present paper, we restrict in Secs. \ref{3}, \ref{4} and \ref{5} to gauge-invariant operators.\par
In gauge-fixed perturbation theory and in renormalization schemes that preserve the BRST invariance, gauge-invariant operators, which after gauge-fixing become BRTS invariant, may mix \cite{JL,He,C} under renormalization only with themselves, with operators that are BRTS exact, and with operators that vanish by the equations of motion. The BRST-exact operators may mix \cite{JL,He,C} only with themselves and with operators that vanish by the equations of motion. The latter may mix \cite{JL,He,C} only with themselves. Thus, the mixing matrix for gauge-invariant operators has a triangular structure \cite{JL,He,C} that simplifies the computation of the anomalous dimensions. \par
For example, $\Tr F^2$ mixes in QCD with certain dimension-$4$ BRST-exact operators \cite{Spiridonov}, and with the gauge-invariant operators, $m \bar\psi \psi$ and $\bar\psi(\slashed{D}+m)\psi$, \footnote{We employ the Euclidean notation.} \cite{Spiridonov}. The latter vanishes by the equations of motion.\par
Now, the correlators of BRST-exact operators with gauge-invariant operators vanish \cite{JL,He,C}. Moreover, the insertion in the vev of operators that vanish by the equations of motion may only produce contact terms \cite{Simma}. Therefore, by limiting ourselves to correlators of gauge-invariant operators at different points, the mixing of gauge-invariant operators with the aforementioned ones may be safely ignored.\par
For computational reasons, we specialize from now on to a QCD-like theory massless in perturbation theory -- a massless QCD-like theory for short --. \par
In a massless QCD-like theory there is no mixing of $\Tr F^2$ with operators containing a mass parameter -- like $m \bar\psi \psi$ -- since they vanish. Besides, according to the above discussion, the mixing of $\Tr F^2$ with the remaining dimension-$4$ operators may be ignored -- up to, perhaps, contact terms -- in gauge-invariant correlators.\par
Therefore, as far as the computations in the present paper are concerned, we may consider $\Tr F^2$ to be multiplicatively renormalizable with a well-defined anomalous dimension. \par
Moreover, restricting to a massless QCD-like theory is specifically convenient because a vast family of canonically normalized gauge-invariant operators exists whose
correlators are conformal invariant to the order of $g^2$ \cite{conformal}. This holds for primary conformal operators to the order $g^2$ \cite{conformal} in the conformal renormalization scheme \cite{conformal}. \par
Indeed, the conformal symmetry is manifest to the order of $g^2$ in a massless QCD-like theory, since the beta function only affects the solution of the Callan-Symanzik (CS) equation starting from the order of $g^4$ \cite{conformal,MBG1}.  \par
Besides, due to the conformal symmetry, provided that the mixing matrix can be diagonalized \footnote{The precise conditions are worked out in \cite{BB}.}, an orthogonal basis of canonically normalized primary conformal operators exists, such that the mixed $2$-point correlators, $\langle O_i(x) O_k(0) \rangle$, vanish for $i \neq k$ to the order of $g^2$, as we show momentarily. \par
For primary conformal operators, $O_i$, with different conformal dimensions to the order of $g^2$, $\Delta_i=D_i-\gamma^{(O_i)}_0 g^2$, with $\gamma_0^{(O_i)}$ the first coefficient of the anomalous dimension defined in Eq. \eqref{gamma}, the orthogonality is a consequence of the conformal symmetry.
For primary conformal operators of spin $s$ with the same conformal dimension, the 2-point mixed correlator in a generic operator basis reads for $x \neq 0$ in the conformal scheme to the order of $g^2$:
\be
G^{(2)}_{ik}(x)=\langle O_i(x)  O_k(0) \rangle  = A_{ik}^{(s)}(g) \, \f{P^{(s)}(x)}{x^{2D}}\left(1 + g^2(\mu) \gamma_0 \log(x^2 \mu^2)\right) 
\ee
with $P^{(s)}(x)$ the spin projector in the coordinate representation in the conformal limit, $A_{ik}^{(s)}(g)$ a constant matrix, $D$ the canonical dimension, and $\gamma_0$ the common first coefficient of the anomalous dimension of the operators. 
Thus, since for gauge-invariant operators that satisfy the spin-statistics theorem the matrix $A_{ik}^{(s)}(g)$ is symmetric, it can always be diagonalized by a change of the operator basis \footnote{For a change of the operator basis, $O'=S O$ in matrix notation, $G^{(2)}$ transforms as $G^{(2)'}= S G^{(2)} S^{T}$, with $S^{T}$ the transposed of the matrix $S$ \cite{BB}. Therefore, a suitable transformation can always diagonalize the -- possibly complex -- symmetric matrix $A^{(s)}(g)$.} \cite{MBG1}. The existence of the orthogonal basis to the order of $g^2$ is employed in Sec. \ref{5}.\par
For the above operators the lhs of Eq. \eqref{LowEneT} is UV log divergent in perturbation theory to the order of $g^2$ because of the nontrivial anomalous dimensions of the operators.\par 
The UV log divergences in the lhs must be reproduced in the rhs by the space-time integration as the operator $\Tr F^2$ gets close to each $O_k$. Hence, we may evaluate the divergent parts in the rhs by the OPE. \par
We specialize now to the case $n=2$ and $O_k=O_1$ for $k=1,2$ (Secs. \ref{3} and \ref{5}). \par
In this case, we show momentarily that the leading perturbative UV log divergences in the rhs of Eq. \eqref{LowEneT} arise from the space-time integration of the following OPE (App. \ref{A}):
\be \label{520}
\Tr F^2(x) O_1(y) \sim C^{(\Tr F^2,O_1)}_{O_1}(x-y) O_1(y) + \cdots
\ee  
where the dots include operators different from $O_1$. \par
Indeed, in order to produce, after the space-time integration, the desired UV log divergence, the coefficient $ C^{(\Tr F^2,O_1)}_{O_1}$ must have canonical dimension $4$.\par
Thus, given that $\Tr F^2$ has canonical dimension $4$, only the contribution of the operators $O_k$ with the same canonical dimension as $O_1$, may lead via the OPE, after the space-time integration in the rhs of Eq. \eqref{LowEneT}, to the desired UV log divergence: 
\bea \label{521}
 \Tr F^2(x) O_i(y) \sim \sum_k C^{(\Tr F^2,O_i)}_{O_k}(x-y) O_k(y) + \cdots
\eea
Higher-dimension operators in the OPE furnish UV finite contributions in the rhs, while lower-dimension operators furnish potential power-like divergences that, however, are absent in the lhs of Eq. \eqref{LowEneT} in dimensional regularization. \par
Moreover, for $n=2$, we show in Sec. \ref{5} that, due to the assumed orthogonality of the operators $O_k$ to the order of $g^2$, the leading contribution in perturbation theory to the $3$-point correlator in the rhs of Eq. \eqref{LowEneT} arising from the coefficients of dimension $4$ occurs from the coefficient of the operator $O_1$ itself. \par
Therefore, by employing the OPE in Eq. \eqref{520}, we should be able to verify the equality of the universal, i.e., scheme-independent, divergent parts in both sides of Eq. \eqref{LowEneT}.\par
Indeed, our computation cannot be exact, but it is affected by finite ambiguities due both to the incomplete OPE and to regularizing in the infrared (IR) the integral in the rhs of Eq. \eqref{LowEneT} in perturbation theory (Sec. \ref{3}). \par

\subsection{Perturbative OPE for $F^2(x)F^2(0)$} \label{32}

In order to verify the above statements, the relevant OPE must be known explicitly. This is the case for the operator $O_1=\Tr F^2$, thanks to the results in \cite{Ch,Z1,Z2} recalled here below.\par
For brevity we define $F^2(x) \equiv 2 \Tr F^2(x)$. The perturbative OPE for $F^2(x)F^2(0)$ in massless QCD reads:
\be
\label{OPEexp}
F^2(x)F^2(0) = C_0^{(S)}(x) \mathbb{I} + C_1^{(S)}(x) F^2(0) + \cdots
\ee  
$C_0^{(S)}(x)$ is the coefficient of the identity operator, $\mathbb{I}$, $C_1^{(S)}(x)$ is the coefficient of the operator $F^2$ itself and the dots stand for other operators that are irrelevant for the purpose of checking the universal log-divergent parts in both sides of Eq. \eqref{LowEneT} according to the above discussion. \par 
For multiplicatively renormalized $F^2$ in massless QCD, 
$C_0^{(S)}$ has been computed perturbatively in the $\overline{MS}$ scheme to three loops, both in the momentum and coordinate representation, in \cite{Ch}:
\bea
\label{C0xstr}
C_0^{(S)}(x) & = &  \f{N^2-1}{x^8}\f{48}{\pi^4} \left(1 + g^2(\mu) (A_{0,1}^{(S)} + 2\beta_0 \log(x^2\mu^2))  \right. \nn \\
&{}&\left. 
+ g^4(\mu) (A_{0,2}^{(S)} + A_{0,3}^{(S)}\log(x^2\mu^2) + 3\beta_0^2 \log^2(x^2\mu^2)) \right) \nn \\
&+& \Delta^2 \delta^{(4)}(x)\dfrac{N^2-1}{4\pi^2}\left(1 + \log(\f{\Lambda^2}{\mu^2}) + g^2(\mu)\left(A_{0,4}^{(S)} + A_{0,5}^{(S)}\log(\f{\Lambda^2}{\mu^2})  \right. \right. \nn \\
&{}& - \left.\left.\beta_0 \log^2(\f{\Lambda^2}{\mu^2})\right) + g^4(\mu)\left(A_{0,6}^{(S)} + A_{0,7}^{(S)}\log(\f{\Lambda^2}{\mu^2}) + A_{0,8}^{(S)}\log^2(\f{\Lambda^2}{\mu^2}) \right. \right. \nn \\
&{}& + \left. \left. \beta_0^2 \log^3(\f{\Lambda^2}{\mu^2 })\right)\right) 
\eea
$C_1^{(S)}$ has been computed in the $\overline{MS}$ scheme, in the momentum representation to two loops in \cite{Ch} and to three loops in \cite{Z1}. We perform in App. \ref{B4} its Fourier transform in the coordinate representation:
\bea
\label{C1xstr}
C_1^{(S)}(x) & = & \f{4\beta_0}{\pi^2x^4}g^2(\mu)\left(1 + g^2(\mu)(A_{1,1}^{(S)} + 2\beta_0\log(x^2\mu^2)) +  \right. \nn \\
&{}&\left. 
+ g^4(\mu)(A_{1,2}^{(S)} + A_{1,3}^{(S)}\log(x^2\mu^2) + 3\beta_0^2\log^2(x^2\mu^2))\right)  \nn \\
& + & \delta^{(4)}(x)\left(4 + g^2(\mu)A_{1,4}^{(S)} + g^4(\mu)\left(A_{1,5}^{(S)} + 4\beta_1\log(\f{\Lambda^2}{\mu^2}) \right) \right. \nn \\
&{}& \left. + g^6(\mu)\left(A_{1,6}^{(S)} + 8\beta_2\log(\f{\Lambda^2}{\mu^2}) - 4\beta_0\beta_1\log^2(\f{\Lambda^2}{\mu^2})\right)\right)
\eea
where: 
\bea
\beta_0 & = & \f{1}{(4\pi)^2}\left(\f{11}{3} - \f{2}{3}\f{\Nf}{N}\right) \\
\beta_1 & = & \f{1}{(4\pi)^4}\left(\f{34}{3} - \f{13}{3}\f{\Nf}{N} +\f{\Nf}{N^3}\right) \\
\beta_2 & = & \f{1}{(4\pi)^6}\left( \f{2857}{54} - \f{1709}{54}\f{\Nf}{N} + \f{56}{27}\f{\Nf^2}{N^2} +\f{187}{36}\f{\Nf}{N^3} - \f{11}{18}\f{\Nf^2}{N^4} + \f{\Nf}{4N^5}\right)
\eea
are the first-three coefficients of the QCD beta function, $\frac{\partial g}{\partial \log \Lambda}=\beta(g) = -\beta_0 g^3 -\beta_1 g^5 -\beta_2 g^7 + \cdots$, and $A_{0,j}^{(S)}$, $A_{1,j}^{(S)}$ are finite coefficients computed in \cite{Ch,Z1}.\par
$\delta^{(4)}(x)$ and $\Delta^2 \delta^{(4)}(x)$ are contact terms in the coordinate representation, i.e., distributions supported at coinciding points. They arise from polynomials in the momentum representation. Interestingly, both finite and divergent contact terms occur in the perturbative OPE \cite{Ch,Z2}. The divergent contact terms require further additive renormalizations \cite{Z2} with respect to the multiplicative renormalization of $F^2$ due to its anomalous dimension. \par 

\subsection{Verifying Eq. \eqref{LowEneT} for $\langle F^2 (z) F^2 (0) \rangle$} \label{34}

By exploiting the OPE in Eq. \eqref{OPEexp}, Eq. \eqref{LowEneT} reads perturbatively:
\be
\label{LETope0} 
4 C_0^{(S)}(z) + 2 g^2\dfrac{\partial C_0^{(S)}(z)}{\partial g^2} \sim C_0^{(S)}(z) \int C_1^{(S)}(x)d^4x
\ee
where the symbol, $\sim$, means -- for the perturbative correlators in the present paper -- equality of the universal divergent parts.\par
We test Eq. \eqref{LETope0} in perturbation theory by extracting the divergent parts in both sides from the bare OPE to the order of $g^2$ in Subsec. \ref{3.1} and to the order of $g^4$ in Subsec. \ref{3.2}. Indeed, the low-energy theorem is derived for the bare coupling in Eqs. \eqref{LowEneT} and \eqref{Intro1}. The perturbative bare coefficients are obtained from the renormalized ones by setting $\mu=\Lambda$ in Eqs. \eqref{C0xstr} and \eqref{C1xstr}. \par
Moreover, we observe  (Subsec. \ref{3.1}) that we can also match the $z$ dependence of the finite part in the lhs of Eq. \eqref{LETope0}, provided that we suitably restrict the IR of the integral in the rhs to the domain $\mathcal{D}^{\Lambda}_{\frac{1}{z}} =  \{ x^2: \frac{1}{\Lambda^2} \leq x^2 \leq z^2 \}$.

\subsection{Verifying Eq. \eqref{LowEneT1} by the RG-improved OPE} \label{35}

We apply the very same IR subtraction prescription to compute the nonperturbative UV asymptotics in both sides of Eq. \eqref{LET20}, which is the analog version of Eq.\eqref{LETope0} that arises from Eqs. \eqref{LowEneT1} and \eqref{LowEneT10}:
\bea
\label{LET2}
&&  \left(\f{\beta(g)}{g}\right)^2  2\Lambda_{QCD}^2 \dfrac{\partial C_0^{(S)}(z)}{\partial \Lambda^2_{QCD}}  \nonumber \\
&&\sim - \left(\f{\beta(g)}{g}\right)^2C_0^{(S)}(z) \int_{\mathcal{D}^{\Lambda}_{\frac{1}{z}}} \f{\beta(g)}{g}C_1^{(S)'}(x)d^4x
\eea
where the symbol, $\sim$, means -- for the RG-improved correlators in the present paper -- asymptotic equality in the UV as $z \rightarrow 0$.\par
Eq. \eqref{LET2} follows from analog arguments of the perturbative case (Subsec. \ref{000}), by taking into account that the leading contribution from the RG-improved OPE is the one that involves the operator $O_1$ itself. \par
Naively, this occurs because the RG-improved version of the OPE is just the resummation of perturbation theory.\par
In fact, the extra factors of $g^2$ that suppress the contributions of other operators in perturbation theory (Subsec. \ref{000}) get transformed into extra factors of the running coupling in the RG-improved OPE. Moreover, the IR subtraction prescription that we employ in the rhs effectively amounts to include only contributions in the OPE that arise as all the coordinates in the rhs are uniformly rescaled. In this situation the above argument implies that the leading contribution in the rhs of Eq. \eqref{LowEneT1} due to the RG-improved OPE only involves the operator $O_1$.\par
Our computation in Sec. \ref{4} involves the nonperturbative UV asymptotics of the OPE coefficients of $F^2$ in the coordinate representation, which we establish a priori, up to a constant overall normalization, by means of the Callan-Symanzik (CS) equation in App. \ref{A}, following standard methods worked out in \cite{MBM,MBN}.
Then, we employ the perturbative results in App. \ref{B} to fix the overall normalization as well. \par
Alternatively, following \cite{MBM} we verify in App. \ref{B3} and \ref{B6} the nonperturbative UV asymptotics of the RG-invariant coefficients, $\left(\f{\beta(g)}{g}\right)^2C_0^{(S)}$ and $- \f{\beta(g)}{g}C_1^{(S)}$, previously derived a priori in App. \ref{A}, by means of a change of the perturbative renormalization scheme that allows us to rewrite the perturbative results, originally expressed in terms of $g(\mu) $ and large logs \cite{Ch,Z1,Z2}, in terms of the running coupling, $g(x)$, to the given perturbative order. \par

\section{Low-energy theorem for $\langle F^2(z) F^2(0) \rangle $ in perturbation theory} \label{3}

Perturbatively, for $n=2$ and $O_k=F^2$, Eq. \eqref{LowEneT} becomes:
\be
\label{PTlet}
4 \langle  F^2(z) F^2(0) \rangle +2 g^2\dfrac{\partial \langle  F^2(z) F^2(0) \rangle}{\partial g^2} = \frac{1}{2}\int \langle  F^2(z) F^2(0) F^2(x)\rangle d^4x
\ee 
since the condensate, $\langle F^2 \rangle$, vanishes identically in dimensional regularization to every order in perturbation theory. \par 
We verify Eq. \eqref{PTlet} in perturbation theory for the bare operator $F^2$. We choose $z \neq 0$ in order to skip the inessential contact terms in the lhs of 
Eq. \eqref{PTlet}. The lhs is log divergent \cite{MBL} to the order of $g^2$ because of the nonvanishing anomalous-dimension coefficient, $\gamma^{(F^2)}_0=2 \beta_0$, of $F^2$, where for a canonically normalized operator, $O$:
\be \label{gamma}
\gamma_O(g) = - \f{\partial\log Z^{(O)}}{\partial\log \mu}=-\gamma^{(O)}_{0}g^2 - \gamma^{(O)}_{1}g^4 + \cdots
\ee
Only $\gamma^{(O)}_0$ is scheme independent in general.
By a standard argument reported in \cite{MBM} the anomalous dimension of $F^2$  \cite{Spiridonov} is related to the beta function in $\overline{MS}$-like schemes:
\be
\gamma_{F^2}(g) = g\f{\partial}{\partial g}\left(\f{\beta(g)}{g}\right)
\ee
since the trace of the stress-energy tensor in massless QCD-like theories is \cite{Spiridonov} RG invariant and proportional to $\frac{\beta(g)}{g} \Tr F^2$ in $\overline{MS}$-like schemes.
It follows:
\bea \label{ad}
\gamma_{F^2}(g) = -2\beta_0 g^2 - 4\beta_1 g^4 + \cdots
\eea
We exhibit in App. \ref{B3} the scheme where manifestly $\gamma^{(F^2)}_1= 4 \beta_1$ following \cite{MBM}. It turns out to be the scheme where the constant finite parts 
of $g^4 C_0^{(S)}(z)$ for $z \neq 0$ vanish to the order of $g^8$.\par
Hence, also the rhs in Eq. \eqref{PTlet} must be divergent, and the divergence can be evaluated by means of the OPE in Eq. \eqref{OPEexp} \cite{MBL}.
Thus, it suffices to evaluate perturbatively the 3-point correlator, $\langle F^2(z)F^2(0)F^2(x)\rangle$, by fixing $z$ while $x$ may be close either to $z$ or $0$. \par
For $x$ close to $0$ we get:
\be
\label{3ptOPE}
\langle F^2(z)F^2(0)F^2(x) \rangle =  C_1^{(S)}(x)\langle F^2(z)F^2(0)\rangle + \dots
\ee 
Therefore, as far as the divergent parts are concerned, the low-energy theorem reads (Subsec. \ref{34}):
\be
\label{AA}
4 C_0^{(S)}(z) + 2 g^2\dfrac{\partial C_0^{(S)}(z)}{\partial g^2} \sim C_0^{(S)}(z) \int C_1^{(S)}(x)d^4x
\ee
where a factor of 2 has been included in the rhs to take into account that $x$ can be close either to $z$ or $0$ \cite{MBL}.\par
To evaluate Eq. \eqref{AA} we employ in the rhs the perturbative version of $C_1^{(S)}$.
After extracting from $C_1^{(S)}$ the lowest-order contact term:
\bea
C_1^{(S)}(x)= 4 \delta^4(x) + C_1^{(S)'}(x)
\eea
Eq. \eqref{AA} simplifies significantly:
\bea
\label{AA1}
2 g^2\dfrac{\partial C_0^{(S)}(z)}{\partial g^2} \sim C_0^{(S)}(z) \int C_1^{(S)'}(x)d^4x
\eea
Interestingly, the lowest-order contact term is crucial to satisfy Eq. \eqref{AA} to the order of $g^0$. In \cite{MBL} both this contact term and the 
compensating first term in the lhs of Eq. \eqref{AA} have been skipped (Subsec. \ref{3.1}).

\subsection{Order of $g^2$} \label{3.1}

To verify Eq. \eqref{AA1} to the order of $g^2$ \cite{MBL}, we employ the corresponding bare OPE coefficients in the coordinate representation. 
They are obtained simply setting $\mu=\Lambda$ in the renormalized ones:
\bea
\label{F2ren}
C_0^{(S)}(z)& = &\dfrac{N^2-1}{z^8}\dfrac{48}{\pi^4}\left(1 + g^2(\Lambda)\left(A_{0,1}^{(S)} +2\beta_0\log(\f{\Lambda^2}{\mu^2}) +2\beta_0\log(z^2\mu^2)\right) \right)
\eea
\bea
\label{C1}
C_1^{(S)'}(x)& = &   \dfrac{1}{x^4}\dfrac{4\beta_0}{\pi^2} g^2(\Lambda)
\eea
In Eq. \eqref{F2ren} we have skipped the inessential contact terms in $C_0^{(S)}(z)$ by choosing $z \neq 0$. 
Hence, the divergent part of Eq. \eqref{AA1} reads to the order of $g^2$:
\be
\label{LET1}
\dfrac{N^2-1}{z^8}\dfrac{48}{\pi^4}g^2(\Lambda)4\beta_0\log(\f{\Lambda^2}{\mu^2}) = \dfrac{N^2-1}{z^8}\dfrac{48}{\pi^4}\frac{4\beta_0}{\pi^2}g^2(\Lambda)\int\frac{1}{x^4}d^4x
\ee
The integral in the rhs is both UV and IR divergent \cite{MBL}.\par
Incidentally, this divergence plays a key role for the compatibility of the open/closed string duality with perturbation theory in massless QCD \cite{MBL}.\par
We regularize the integral by restricting to the domain $\mathcal{D}^{\Lambda}_{\mu} =  \{ x^2: \frac{1}{\Lambda^2} \leq x^2 \leq \frac{1}{\mu^2} \} $. Hence, performing the integral in polar coordinates, we get: 
\be
\int_{\mathcal{D}^{\Lambda}_{\mu}} \dfrac{1}{x^4}d^4x = \pi^2\log(\frac{\Lambda^2}{\mu^2})
\ee
that implies Eq. \eqref{LET1} according to \cite{MBL}.\par
Our key observation is that we can also match the $z$ dependence of the finite part in the lhs of Eq. \eqref{AA1}, provided that we suitably modify the integration domain in the IR, $\mathcal{D}^{\Lambda}_{\frac{1}{z}} =  \{ x^2: \frac{1}{\Lambda^2} \leq x^2 \leq z^2 \} $. By including the constant finite parts for future employment, we get to the order of $g^2$:
\bea
&&\dfrac{N^2-1}{z^8}\dfrac{48}{\pi^4}g^2(\Lambda) \left(2A_{0,1}^{(S)} + 4\beta_0\log(z^2\Lambda^2) \right) \nonumber \\
&&= \dfrac{N^2-1}{z^8}\dfrac{48}{\pi^4} \left(A_{1,4}^{(S)} + \frac{4\beta_0}{\pi^2} \int_{\mathcal{D}^{\Lambda}_{\frac{1}{z}} }\frac{1}{x^4}d^4x \right) g^2(\Lambda)
\eea
where now the equality also includes the $z$ dependence up to the constant finite parts.
This prescription plays a key role for getting the correct nonperturbative RG-improved UV asymptotics in the rhs of Eq. \eqref{LET2}. \par
Of course, any prescription for the IR cutoff in the rhs leads, already to the order of $g^2$, to an ambiguity for the constant finite parts in the rhs. \par
Presently, we cannot resolve this finite ambiguity in the framework of our computation that is either based in this Sec. on the perturbative OPE or in Sec. \ref{4} on its universal nonperturbative asymptotics.

\subsection{Order of $g^4$} \label{3.2}

The $O(g^2)$ UV log divergence computed in Subsec. \ref{3.1} is universal, i.e., it depends only on the first coefficient of the anomalous dimension.
The $O(g^4)$ log-squared divergence in the lhs of Eq. \eqref{AA1} is universal as well, because it is essentially the square of the $O(g^2)$ log divergence. \par
Instead, the $O(g^4)$ log divergence is scheme dependent, and therefore depends on the constant finite parts to the order of $g^2$. 
Thus, we verify perturbatively Eq. \eqref{AA1} by limiting ourselves to the universal divergences. \par
However, we compute as well the scheme-dependent finite and log-divergent parts in both sides of Eq. \eqref{AA1} to the order of $g^4$ for future employment.
We evaluate the rhs of Eq. \ref{AA1} to the order of $g^4$:
\bea
&&  \f{N^2-1}{z^8}\f{48}{\pi^4} \left(1 + g^2(\Lambda)( A_{0,1}^{(S)} +2\beta_0\log(z^2\Lambda^2) ) \right)  \nonumber \\
&& g^2(\Lambda)  \int_{\mathcal{D}^{\Lambda}_{\frac{1}{z}} } \f{4\beta_0}{\pi^2x^4} \left(1 + g^2(\Lambda)(A_{1,1}^{(S)} + 2 \beta_0 \log(x^2\Lambda^2)) \right) \nn \\
&& +  \delta^{(4)}(x)\left( A_{1,4}^{(S)} + g^2(\Lambda)(A_{1,5}^{(S)} + 4\beta_1\log(\f{\Lambda^2}{\mu^2})) \right)
d^4x 
\eea
It reads to the order of $g^4$:
\bea
&& \f{N^2-1}{z^8}\f{48}{\pi^4} g^4(\Lambda) \left( \int_{\mathcal{D}^{\Lambda}_{\frac{1}{z}} } \f{4\beta_0}{\pi^2x^4} \left(A_{1,1}^{(S)} + 2 \beta_0 \log(x^2\Lambda^2) \right) 
d^4x + A_{1,5}^{(S)} \right. \nonumber \\
&& \left. + 4\beta_1\log(\f{\Lambda^2}{\mu^2}) \right) + \f{N^2-1}{z^8}\f{48}{\pi^4}  g^4(\Lambda) \left( A_{0,1}^{(S)} +2\beta_0\log(z^2\Lambda^2) \right) \int_{\mathcal{D}^{\Lambda}_{\frac{1}{z}} } \f{4\beta_0}{\pi^2x^4} d^4x \nn \\
&& + \f{N^2-1}{z^8}\f{48}{\pi^4}  g^4(\Lambda) \left( A_{0,1}^{(S)} +2\beta_0\log(z^2\Lambda^2) \right)A_{1,4}^{(S)}
\eea
Hence, we evaluate Eq. \ref{AA1} to the order of $g^4$:
\bea
&&\f{N^2-1}{z^8}\f{48}{\pi^4}g^4(\Lambda) \left(4A_{0,2}^{(S)} + 4A_{0,3}^{(S)}\log(z^2\Lambda^2) + 12\beta_0^2\log^2(z^2\Lambda^2)\right) \nonumber \\
&&  \sim \f{N^2-1}{z^8}\f{48}{\pi^4} g^4(\Lambda) \left( \int_{\mathcal{D}^{\Lambda}_{\frac{1}{z}} } \f{4\beta_0}{\pi^2x^4} \left(A_{1,1}^{(S)} + 2 \beta_0 \log(x^2\Lambda^2) \right) 
d^4x + A_{1,5}^{(S)} \right. \nonumber \\
&& \left. \;\;\;\; + 4\beta_1\log(\f{\Lambda^2}{\mu^2}) \right) + \f{N^2-1}{z^8}\f{48}{\pi^4}  g^4(\Lambda) \left( A_{0,1}^{(S)} +2\beta_0\log(z^2\Lambda^2) \right) \int_{\mathcal{D}^{\Lambda}_{\frac{1}{z}} } \f{4\beta_0}{\pi^2x^4} d^4x \nn \\
&& \;\;\;\;\,+\f{N^2-1}{z^8}\f{48}{\pi^4}  g^4(\Lambda) \left( A_{0,1}^{(S)} +2\beta_0\log(z^2\Lambda^2) \right)A_{1,4}^{(S)} \nn \\
&&  = \f{N^2-1}{z^8}\f{48}{\pi^4} g^4(\Lambda) \left(4\beta_0 A_{1,1}^{(S)} \log(z^2\Lambda^2) + 4 \beta_0^2 \log^2(z^2\Lambda^2) 
 + A_{1,5}^{(S)} \right. \nonumber \\
&& \left. \;\;\;\;+ 4\beta_1\log(\f{\Lambda^2}{\mu^2}) \right) + \f{N^2-1}{z^8}\f{48}{\pi^4}  g^4(\Lambda) \left( A_{0,1}^{(S)} +2\beta_0\log(z^2\Lambda^2) \right) 4\beta_0 \log(z^2\Lambda^2) \nonumber \\
&& \;\;\;\;\,+ \f{N^2-1}{z^8}\f{48}{\pi^4}  g^4(\Lambda) \left( A_{0,1}^{(S)} +2\beta_0\log(z^2\Lambda^2) \right) A_{1,4}^{(S)} \nn \\
&&  = \f{N^2-1}{z^8}\f{48}{\pi^4} g^4(\Lambda) \left(4\beta_0 A_{1,1}^{(S)} \log(z^2\Lambda^2) + 4 \beta_0^2 \log^2(z^2\Lambda^2) 
 + A_{1,5}^{(S)} \right. \nonumber \\
&& \left. \;\;\;\;+ 4\beta_1\log(\f{\Lambda^2}{\mu^2}) \right) + \f{N^2-1}{z^8}\f{48}{\pi^4}  g^4(\Lambda) \left( A_{0,1}^{(S)} A_{1,4}^{(S)} + 8 \beta_0^2  \log^2(z^2\Lambda^2) \right. \nonumber \\
&& \left.  \;\;\;\;+ 2 \beta_0 A_{1,4}^{(S)} \log(z^2\Lambda^2) + 4 \beta_0 A_{0,1}^{(S)} \log(z^2\Lambda^2)\right) \nn \\
&&  = \f{N^2-1}{z^8}\f{48}{\pi^4} g^4(\Lambda)  \left( A_{1,5}^{(S)} + A_{0,1}^{(S)} A_{1,4}^{(S)}  + 4\beta_0 (A_{1,1}^{(S)}+ A_{0,1}^{(S)}+ \frac{1}{2} A_{1,4}^{(S)}) \log(z^2\Lambda^2)  
 \right. \nonumber \\
&& \left.
\;\;\;\;+ 4\beta_1\log(\f{\Lambda^2}{\mu^2}) + 12 \beta_0^2  \log^2(z^2\Lambda^2) \right)
\eea
where we recall that the symbol, $\sim$, means -- for the perturbative correlators in the present paper -- equality of the universal divergent parts.\par
Therefore, the universal log-squared divergences in both sides of Eq. \eqref{AA1} agree. 

\section{Nonperturbative UV asymptotics of the low-energy theorem for $\langle F^2(z) F^2(0) \rangle $} \label{4}

We compute now the nonperturbative UV asymptotics of the low-energy theorem for $\langle F^2(z) F^2(0) \rangle$ in massless QCD, within the universal leading and next to leading logarithmic accuracy, by means of the UV asymptotics of the renormalized OPE coefficients in App. \ref{A}, and of their perturbative normalization in App. \ref{B}. \par
It is convenient to introduce the RG-invariant coefficients $\left(\f{\beta(g)}{g}\right)^2C_{0}^{(S)}(z) $ and $-\f{\beta(g)}{g}C_1^{(S)}(x)$ associated to the OPE of the RG-invariant
operator $-\f{\beta(g)}{g} F^2$. \par The change of normalization does not affect the universal UV asymptotics but for the overall normalization, yet it is specifically convenient for the perturbative computations in App. \ref{B}:
the universal UV asymptotics of the 2-point correlators of $ -\f{\beta(g)}{g} F^2$ that is RG invariant, of $F^2$ that has a nontrivial anomalous dimension, and of $g^2 F^2$ whose first coefficient of the anomalous dimension vanishes (App. \ref{B3}), coincide up to the overall normalization \cite{MBM}. \par
The UV universal asymptotics of the RG-invariant OPE coefficients reads:
\bea \label{F2UV}
\left(\f{\beta(g)}{g}\right)^2C_0^{(S)}(z) &\sim& \dfrac{N^2-1}{z^8}\dfrac{48}{\pi^4}\beta_0^2g^4(z) \nonumber \\
&\sim& \dfrac{N^2-1}{\pi^4}\dfrac{48}{z^8}\dfrac{1}{\log^2(\frac{1}{z^2\Lambda^2_{QCD}})}\left(1-2\frac{\beta_1}{\beta_0^2}\frac{\log\log(\frac{1}{z^2\Lambda^2_{QCD}})}{\log(\frac{1}{z^2\Lambda^2_{QCD}})}\right)
\eea
and:
\bea
\label{C1UV}
- \f{\beta(g)}{g}C_1^{(S)'}(x) &\sim& \dfrac{4\beta_0^2}{\pi^2}\dfrac{1}{x^4}g^4(x) \nonumber \\
&\sim& \dfrac{4}{\pi^2}\dfrac{1}{x^4}\dfrac{1}{ \log^2(\frac{1}{x^2\Lambda^2_{QCD}})}\left(1-2\frac{\beta_1}{\beta_0^2}\frac{\log\log(\frac{1}{x^2\Lambda^2_{QCD}})}{\log(\frac{1}{x^2\Lambda^2_{QCD}})}\right)
\eea
where we recall that the symbol, $\sim$, means -- for the RG-improved correlators in the present paper -- asymptotic equality in the UV as $z,x \rightarrow 0$. The asymptotic equalities in the second lines of Eqs. \eqref{F2UV} and \eqref{C1UV} 
follow from Eq. \eqref{alfa}.\par
It is convenient to employ the version of the low-energy theorem that involves $\Lambda_{QCD}$ and the canonical normalization of the YM action in Eq. \eqref{LowEneT1}. We skip the finite contact term in $C_1^{(S)}$, the compensating term in the lhs of Eq. \eqref{LowEneT1}, and the divergent contact terms in $C_1^{(S)}$ that, according to \cite{Z2}, should be renormalized to zero. \par
Then, for the renormalized correlators, it should hold nonperturbatively and asymptotically as $z \rightarrow 0$ (Subsec. \ref{35}):
\bea
\label{LET20}
&&  \left(\f{\beta(g)}{g}\right)^2  2\Lambda_{QCD}^2 \dfrac{\partial C_0^{(S)}(z)}{\partial \Lambda^2_{QCD}}  \nonumber \\
&&\sim - \left(\f{\beta(g)}{g}\right)^2C_0^{(S)}(z) \int_{\mathcal{D}^{\Lambda}_{\frac{1}{z}}} \f{\beta(g)}{g}C_1^{(S)'}(x)d^4x
\eea
Firstly, we compute the lhs of Eq. \eqref{LET20}: 
\bea
\label{lhs1}
&&     \left(\f{\beta(g)}{g}\right)^2  2\Lambda_{QCD}^2 \dfrac{\partial C_0^{(S)}(z) }{\partial \Lambda^2_{QCD}} \nonumber \\
&& \sim  \dfrac{N^2-1}{\pi^4}\dfrac{48}{z^8}\dfrac{4}{\log^3(\frac{1}{z^2\Lambda^2_{QCD}})}\left(1-3\frac{\beta_1}{\beta_0^2}\frac{\log\log(\frac{1}{z^2\Lambda^2_{QCD}})}{\log(\frac{1}{z^2\Lambda^2_{QCD}})}\right) \nonumber \\
&&\sim  \dfrac{N^2-1}{\pi^4}\dfrac{48}{z^8}4\beta_0^3g^6(z)
\eea
In the rhs of Eq. \eqref{LET20} the crucial step is the integration of $-\f{\beta(g)}{g}C_1^{(S)}$. According to the prescription in Subsec. \ref{3.1},
we have restricted the integral to the domain $\mathcal{D}^{\Lambda}_{\frac{1}{z}} =  \{ x^2: \frac{1}{\Lambda^2} \leq x^2 \leq z^2 \} $.\par
But now, after the RG resummation, the integral is UV convergent because of the asymptotic freedom, and therefore we can remove the UV cutoff. Thus, we may extend the integration to the new domain $\mathcal{D}^{\infty}_{\frac{1}{z}} =  \{ x^2: 0 \leq x^2 \leq z^2 \} $. \par
Incidentally, the nonperturbative UV finiteness of the integral in Eq. \eqref{LET20}, as opposed to the UV divergence of the integral in Eq. \eqref{LET1} in perturbation theory, plays a key role for the no-go theorem in \cite{MBL}.
We get:
\bea
\label{C1int}
&& - \int_{\mathcal{D}^{\infty}_{\frac{1}{z}} } \f{\beta(g)}{g}C_1^{(S)'}(x)d^4x \nonumber \\
&& \sim  \int_{\mathcal{D}^{\infty}_{\frac{1}{z}}} \dfrac{4}{\pi^2 x^4}\dfrac{1}{\log^2(\frac{1}{x^2\Lambda^2_{QCD}})}\left(1-2\frac{\beta_1}{\beta_0^2}\frac{\log\log(\frac{1}{x^2\Lambda^2_{QCD}})}{\log(\frac{1}{x^2\Lambda^2_{QCD}})}\right)d^4x  \nonumber \\
&& \sim  4\dfrac{1}{\log(\frac{1}{z^2\Lambda^2_{QCD}})}\left(1- \frac{\beta_1}{\beta_0^2}\frac{\log\log(\frac{1}{z^2\Lambda^2_{QCD}})}{\log(\frac{1}{z^2\Lambda^2_{QCD}})}\right) 
\nonumber \\
&& \sim 4 \beta_0 g^2(z)
\eea
that substituted in the rhs of Eq. \eqref{LET2} implies:
\bea
&&- \left(\f{\beta(g)}{g}\right)^2C_0^{(S)}(z) \int_{\mathcal{D}^{\infty}_{\frac{1}{z}}} \f{\beta(g)}{g}C^{(S)'}_1(x)d^4x \nonumber \\
&& \sim \dfrac{N^2-1}{z^8}\dfrac{48}{\pi^4}\beta_0^2g^4(z)4 \beta_0 g^2(z) \nonumber \\
&& \sim  \dfrac{N^2-1}{z^8}\dfrac{48}{\pi^4}4\beta_0^3 g^6(z)
\eea
that actually matches Eq. \eqref{lhs1}.\par
Just as an aside, the integral in Eq. \eqref{C1int} is computed in polar coordinates by means of the obvious change of variables and by integrating by parts:
\bea
&& \int_{\mathcal{D}^{\infty}_{\frac{1}{z}} } \dfrac{1}{\pi^2 x^4}\dfrac{1}{\log^2(\frac{1}{x^2\Lambda^2_{QCD}})}\left(1-2\frac{\beta_1}{\beta_0^2}\frac{\log\log(\frac{1}{x^2\Lambda^2_{QCD}})}{\log(\frac{1}{x^2\Lambda^2_{QCD}})}\right)d^4x \nn \\
&& = \int^{|z|}_0  \dfrac{2}{\log^2(\frac{1}{|x|^2\Lambda^2_{QCD}})}\left(1-2\frac{\beta_1}{\beta_0^2}\frac{\log\log(\frac{1}{|x|^2\Lambda^2_{QCD}})}{\log(\frac{1}{|x|^2\Lambda^2_{QCD}})}\right) \frac{d |x| }{|x|} \nn \\
&& = \f{1}{\log\f{1}{z^2\Lambda^2_{QCD}}} - \f{\beta_1}{\beta_0^2}\f{\log\log\f{1}{z^2\Lambda^2_{QCD}}}{\log^2\f{1}{z^2\Lambda^2_{QCD}}}-\f{\beta_1}{2\beta_0^2}\f{1}{\log^2\f{1}{z^2\Lambda^2_{QCD}}} \nn \\
&& \sim \dfrac{1}{\log(\frac{1}{z^2\Lambda^2_{QCD}})}\left(1- \frac{\beta_1}{\beta_0^2}\frac{\log\log(\frac{1}{z^2\Lambda^2_{QCD}})}{\log(\frac{1}{z^2\Lambda^2_{QCD}})}\right) 
\eea
with $|x|=\sqrt{x^2}$.

\section{Perturbative OPE to the order of $g^2$ and nonperturbative UV asymptotics from the low-energy theorem for $\langle O(z) O(0) \rangle$} \label{5}

By inverting the arguments in the preceding Secs., we now employ the low-energy theorem in order to get information on the OPE coefficients. \par
We consider, in a perturbatively massless QCD-like theory, the operators, $O_i$, mentioned in Subsec. \ref{000} that we assume to be gauge invariant,
to have the same canonical dimension, and to mix under renormalization. Then, for $n=2$ Eq. \eqref{LowEneT0} reads:
\bea 
&&2 \big(c_i + c_k)  \langle O_i(z) O_k(0) \rangle + 2 \dfrac{\partial \langle O_i(z) O_k(0) \rangle}{\partial \log g} \nonumber \\
&&= \int \langle O_i(z) O_k(0)  F^2(x) \rangle - \langle O_i(z) O_k(0) \rangle \langle F^2(x) \rangle d^4x
\eea
By exploiting the OPE in Eq. \eqref{521}, we get perturbatively:
\bea \label{o}
&&2 \big(c_i + c_k)  \langle O_i(z) O_k(0) \rangle + 2 \dfrac{\partial \langle O_i(z) O_k(0) \rangle}{\partial \log g} \nonumber \\
&& \sim  \langle O_i(z) O_k(0) \rangle \int  C^{(F^2,O_i)}_{O_i}(x-z) +  C^{(F^2,O_k)}_{O_k}(x) d^4x \nonumber \\
&& + \sum_{l \neq i,k} \langle O_l(z) O_k(0) \rangle \int  C^{(F^2,O_i)}_{O_l}(x-z)  d^4x \nonumber \\
&& + \sum_{l \neq k,i} \langle O_i(z) O_l(0) \rangle \int  C^{(F^2,O_k)}_{O_l}(x)  d^4x
\eea
As the operators $O_i$ are canonically normalized, only the first term in the lhs of Eq. \eqref{o} contributes perturbatively to the order of $g^0$.
Moreover, to this order, we get:
\bea
2 c_i   \langle O_i(z) O_i(0) \rangle  =  \langle O_i(z) O_i(0) \rangle \int  C^{(F^2,O_i)}_{O_i}(x) d^4x 
\eea
since, because of the orthogonality of the operators with themselves and with the operators with different canonical dimensions, all the remaining contributions in the rhs from the complete OPE vanish to this order.\par
Hence, as the lhs is finite, a finite contact term must occur, for $c_i \neq 0$, to the order of $g^0$.
Indeed, by dimensional analysis, since $F^2$ has canonical dimension $4$:
\bea \label{c}
C_{O_i}^{(F^2,O_i)}(x) =  2 \, c_i \, \delta^4(x) + C_{O_i}^{(F^2,O_i)'}(x) 
\eea
with $C_{O_i}^{(O_i,F^2)'}$ necessarily on the order of $g^2$ -- for nontrivial anomalous dimensions of the operators $O_i$ -- because, by dimensional analysis and the orthogonality of the operators to the order of $g^2$, it has to produce in the rhs the log divergence that matches the one in the lhs, which may only arise to the order of $g^2$ because of the log derivative in the lhs:
\bea \label{c10}
&&  \dfrac{\partial \langle O_i(z) O_i(0) \rangle}{\partial \log g}  \sim  \langle O_i(z) O_i(0) \rangle \int  C^{(F^2,O_i)'}_{O_i}(x)  d^4x 
\eea
Thus, we have reduced our computation to the case of a single multiplicatively renomalizable operator with well-defined anomalous dimension, which we summarize as follows.\par
The low-energy theorem for any 2-point correlator, $\langle O(z) O(0) \rangle $, of a canonically normalized gauge-invariant operator, $O$, belonging to the aforementioned orthogonal basis, implies perturbatively:
\be 
\label{LETope}
2 \, c \, C_0^{(O)}(z) + \dfrac{\partial C_0^{(O)}(z)}{\partial \log g} \sim C_0^{(O)}(z) \int C^{(F^2,O)}_{O}(x) d^4x
\ee
with $c$ the exponent, defined in Subsec. \ref{2.3}, of the canonical rescaling of the operator $O$, and the OPE coefficients defined in Eqs. \eqref{51} and \eqref{52}.
For an operator of spin $s$, the bare $C_0^{(O)}$, up to a finite term on the order of $g^2$, reads to the order of $g^2$ for $z \neq 0$:
\be
C_0^{(O)}(z) = A^{(s)} \, \f{P^{(s)}(z)}{z^{2D}}\left(1 + g^2(\Lambda) \gamma_0^{(O)} \log( z^2 \Lambda^2)\right)
\ee
with $P^{(s)}(z)$ the spin projector in the coordinate representation in the conformal limit, $A^{(s)}$ a constant normalization factor, $D$ the canonical dimension, and $\gamma_0^{(O)}$ the first coefficient of the anomalous dimension. \par 
The first term in the lhs of Eq. \eqref{LETope} implies that a finite contact term must occur for $c \neq 0$ to the order of $g^0$:
\bea 
C_{O}^{(F^2,O)}(x) =  2 \, c \, \delta^4(x) + C_{O}^{(F^2,O)'}(x) 
\eea
By skipping the contact term in the rhs and the compensating term in the lhs of Eq. \eqref{LETope}, Eq. \eqref{c10} implies to the order of $g^2$:
\bea \label{5.4}
\f{2A^{(s)} \,\gamma_0^{(O)} P^{(s)}(z)}{z^{2D}} g^2(\Lambda) \log (z^2\Lambda^2) \sim \f{A^{(s)} \, P^{(s)}(z)}{z^{2D}}\int_{{\mathcal{D}^{\Lambda}_{\frac{1}{z}}}}C_{O}^{(F^2,O)'}(x)d^4x 
\eea
This fixes $C_{O}^{(F^2,O)}$ in a massless QCD-like theory to the order of $g^2$ in terms of the first coefficient of the anomalous dimension, $\gamma_{0}^{(O)}$, and of the exponent, $c$, of the canonical rescaling:
\bea
\label{d}
C_{O}^{(F^2,O)}(x) = 2 \, c \, \delta^4(x) +
g^2(\Lambda) \f{2\gamma_{0}^{(O)}}{\pi^2 x^4}
\eea
Moreover, it follows from Eq. \eqref{as}:
\bea
C_{O}^{(F^2,O)'}(x) & \sim & \f{2\gamma_{0}^{(O)}}{\pi^2 x^{4}} g^{2}(x)\left(\f{g(x)}{g(\mu)}\right)^{2}  \nn \\
&\sim & \f{2\gamma_0^{(O)}}{\pi^2 x^4} \dfrac{1}{g^2(\mu)\beta_0^2 \log^2(\frac{1}{x^2\Lambda^2_{QCD}})}\left(1-2\frac{\beta_1}{\beta_0^2}\frac{\log\log(\frac{1}{x^2\Lambda^2_{QCD}})}{\log(\frac{1}{x^2\Lambda^2_{QCD}})}\right)
\eea
Then, the low-energy theorem implies asymptotically as $z \rightarrow 0$:
\be
\label{5A}
2\Lambda^2_{QCD} \f{\partial C_0^{(O)}(z)}{\partial \Lambda^2_{QCD}} \sim -\f{\beta(g)}{g} C_0^{(O)}(z)\int_{{\mathcal{D}^{\Lambda}_{\frac{1}{z}}}} C_{O}^{(F^2,O)'}(x) d^4x
\ee
with:
\bea
C_0^{(O)}(z) &\sim& A^{(s)} \f{P^{(s)}(z)}{z^{2D}} \left(\f{g(z)}{g(\mu)}\right)^{\f{2 \gamma^{(O)}_{0}}{\beta_0}} \nn \\
&\sim&  \f{1}{z^{2D}}\dfrac{A^{(s)} \, P^{(s)}(z)}{(g^2(\mu))^{\f{\gamma^{(O)}_{0}}{\beta_0}}\beta_0^{\f{ \gamma^{(O)}_{0}}{\beta_0}} \log^{\f{ \gamma^{(O)}_{0}}{\beta_0}}(\frac{1}{z^2\Lambda^2_{QCD}})}\left(1-\f{ \gamma^{(O)}_{0}}{\beta_0}\frac{\beta_1}{\beta_0^2}\frac{\log\log(\frac{1}{z^2\Lambda^2_{QCD}})}{\log(\frac{1}{z^2\Lambda^2_{QCD}})}\right) \nn \\
\eea
The lhs of Eq. \eqref{5A} reads:
\begin{small}
\bea
\label{5A1}
&&2\Lambda^2_{QCD} \f{\partial C_0^{(O)}(z)}{\partial \Lambda^2_{QCD}} \nn \\
&& \sim  \f{2\gamma_{0}^{(O)}}{z^{2D}}\f{A^{(s)} \, P^{(s)}(z)}{(g^2(\mu))^{\f{\gamma_{0}^{(O)}}{\beta_0}} \beta_0^{1+\f{\gamma_{0}^{(O)}}{\beta_0}}\log^{1+\f{\gamma_{0}^{(O)}}{\beta_0}}(\f{1}{z^2\Lambda^2_{QCD}})}\left(1 - (1+\f{\gamma_{0}^{(O)}}{\beta_0})\f{\beta_1}{\beta_0^2}\f{\log\log(\f{1}{z^2\Lambda^2_{QCD}})}{\log(\f{1}{z^2\Lambda^2_{QCD}})}\right) \nn \\
&& \sim A^{(s)} \, P^{(s)}(z) \f{2\gamma_{0}^{(O)}}{z^{2D}}(\frac{g(z)}{g(\mu)})^{\f{2\gamma_{0}^{(O)}}{\beta_0}} g^2(z)
\eea
\end{small}
The integral in the rhs of Eq. \eqref{5A} is UV convergent exactly as in Sec. \ref{4}, and the integration domain can be extended to $\mathcal{D}^{\infty}_{\frac{1}{z}}$:
\bea
&& -\f{\beta(g)}{g} \int_{\mathcal{D}^{\infty}_{\frac{1}{z}}} C_{O}^{(F^2,O)'}(x) d^4x   \\
&& \sim  \int_{\mathcal{D}^{\infty}_{\frac{1}{z}}} \f{2\gamma_{0}^{(O)}\beta_0}{\pi^2x^4}\dfrac{1}{\beta_0^2\log^2(\frac{1}{x^2\Lambda^2_{QCD}})}\left(1-2\frac{\beta_1}{\beta_0^2}\frac{\log\log(\frac{1}{x^2\Lambda^2_{QCD}})}{\log(\frac{1}{x^2\Lambda^2_{QCD}})}\right)d^4x  \nn \\
&& \sim  \dfrac{2\gamma_{0}^{(O)}}{\beta_0\log(\frac{1}{z^2\Lambda^2_{QCD}})}\left(1-\frac{\beta_1}{\beta_0^2}\frac{\log\log(\frac{1}{z^2\Lambda^2_{QCD}})}{\log(\frac{1}{z^2\Lambda^2_{QCD}})}\right) \sim 2\gamma_{0}^{(O)} g^2(z)
\eea
Therefore, the rhs of Eq. \eqref{5A} reads:
\bea
&& -\f{\beta(g)}{g} C_0^{(O)}(z)\int_{\mathcal{D}^{\infty}_{\frac{1}{z}}}C_{O}^{(F^2,O)'}(x)d^4x \sim \f{A^{(s)} \, P^{(s)}(z)}{z^{2D}}(\frac{g(z)}{g(\mu)})^{\f{2\gamma_{0}^{(O)}}{\beta_0}}2\gamma_{0}^{(O)}g^2(z)
\eea
that actually matches Eq. \eqref{5A1}.

\section{Conclusions} \label{7}

\subsection{Low-energy theorem and perturbation theory}

As expected, our computations have actually verified the equality to the order of $g^4$ of the perturbative universal divergent parts in both sides of Eq. \eqref{LowEneT}, for $n=2$ and $O_k= F^2$ (Sec. \ref{3}), by means of the OPE of the operator $F^2$ with itself worked out in \cite{Ch,Z1,Z2}. \par
The contact terms in $C_0^{(S)}$ have been, in fact, inessential to verify the low-energy theorem, since they can be skipped by choosing $z \neq 0$ in Eq. \eqref{LETope0}. Instead, the contact terms in $C_1^{(S)}$ could not be skipped, because $C_1^{(S)}$ is integrated over the whole space-time.\par
We have demonstrated that the finite contact term to the order of $g^0$ in $C_1^{(S)}$ in the rhs of Eq. \eqref{LETope0} is crucial to match the first term in the lhs perturbatively (Sec. \ref{3}). Thus, somehow surprisingly, the rationale for its occurrence is the low-energy theorem. \par
We have also observed that the $O(g^4)$ log-divergent contact term in $C_1^{(S)}$ \cite{Ch,Z1,Z2} mixes in Eq. \eqref{LETope0} with the scheme-dependent divergences to the order of $g^4$ due to the second coefficient of the anomalous dimension of $F^2$ (Sec. \ref{3}). Since the latter divergences are affected by the finite ambiguities to the order of $g^2$ (Subsec. \ref{000}), presently we cannot argue on the basis of the low-energy theorem about the renormalization of the $O(g^4)$ log-divergent contact term in $C_1^{(S)}$ perturbatively. \par
We have anyway computed the constant finite parts to the order of $g^4$ in both sides of Eq. \eqref{LETope0}, with the IR subtraction point of the integral in the rhs specified in Sec. \ref{3}, for future employment as well. \par
Indeed, we have found in Subsec. \ref{3.1} the prescription for the IR subtraction point of the integral in the rhs of Eq. \eqref{LETope0} that also allows us to reproduce the finite small-$z$ dependence in the lhs in perturbation theory to the order of $g^2$.  \par
Finally, in a massless QCD-like theory, by inverting the logic of the computation we have derived in Sec. \ref{5} from the low-energy theorem for the 2-point correlator, $\langle O(x) O(0) \rangle$, of a multiplicatively renormalizable gauge-invariant operator, $O$, the perturbative OPE coefficient, $C_{O}^{(F^2,O)}$, to the order of $g^2$ \footnote{After the present paper has been posted in arXiv we have become aware of \cite{OPE1}, where, interestingly, it has been shown how to obtain in principle the perturbative OPE coefficients directly in their renormalized form on the basis of an already renormalized analog -- in the $\phi^4$ theory -- of Eq. \eqref{LET} that has been derived independently of the present paper in the framework of a suitable BPHZ-like renormalization scheme, which does not employ explicitly the functional integral. Yet, the computations are very hard in the aforementioned scheme, and presently no explicit computation has been performed in YM theories \cite{OPE2}, which a renormalized analog of Eq. \eqref{LET} applies to as well \cite{OPE2}.This is to be contrasted with the approach in the present paper, where Eq. \eqref{LET} holds for the bare correlators and allows us an \emph{a priori} evaluation of the log divergences from the anomalous dimensions, so that the OPE coefficients that produce the log divergences are easily evaluated. Moreover, the nonperturbative version of the low-energy theorem in terms of $\Lambda_{QCD}$ in Eq. \eqref{LET0} appears to be new. We would like to thank M. B. Frob for pointing out to us \cite{OPE1,OPE2} and for several discussions about his and our work.}.

\subsection{Low-energy theorem and nonperturbative UV asymptotics}

The aforementioned IR prescription, which also reproduces the $z$ dependence of the finite parts, has played a crucial role to verify the low-energy theorem nonperturbatively asymptotically in the UV.\par
Indeed, by skipping the divergent contact term in $C_1^{(S)}$, but not the finite one, we have verified the low-energy theorem (Sec. \ref{4}) -- Eq. \eqref{LET0} for $n=2$ and $\mathcal{O}_k=F^2$--  nonperturbatively in its canonical version -- Eq. \eqref{LowEneT1} -- by means of the renormalized RG-improved OPE asymptotically in the UV (Apps. \ref{A} and  \ref{B}). This is compatible with the previous result in \cite{Z2}, obtained independently of the low-energy theorem, that the divergent contact term in $C_1^{(S)}$ should be renormalized to zero. \par
Incidentally, for future applications, by working out the nonperturbative UV asymptotics of the OPE for $F^2$, we have computed the scale-dependent corrections to the universal asymptotics of the 2-point correlator of $\frac{g^2}{2N} F^2$ (App. \ref{B3}) - the YM Lagrangian density with the canonical normalization (Subsec. \ref{2.3}), whose correlators coincide with the correlators of the Wilson plaquette on the lattice in the continuum limit (Subsec. \ref{2.3}) -- which is not RG invariant.\par
We have also computed the universal UV asymptotics of the 2-point correlator of its RG-invariant version, $\left(\f{\beta(g)}{g}\right) F^2$ (Sec. \ref{4} and App. \ref{B3})
that coincides (Sec. \ref{4}), but for the overall scale-dependent normalization, with the UV asymptotics of the 2-point correlators of $\frac{g^2}{2} F^2$ and $F^2$ (App. \ref{B3}).\par
Finally, by inverting the logic of the computation, we have derived from the low-energy theorem for any 2-point correlator, $\langle O(x)  O(0) \rangle$, of a multiplicatively renormalizable gauge-invariant operator, $O$, the nonperturbative UV asymptotics in Sec. \ref{5} by means of the lower-order perturbative OPE discussed above. \par
Since the rhs of the low-energy theorem contains an operator insertion at zero momentum, it is somehow surprising that the low-energy theorem is asymptotically verified nonperturbatively in the UV. \par
Yet, our RG-improved computations in Secs. \ref{4} and \ref{5} seem to show \emph{a posteriori} that the integral on space-time in rhs of the low-energy theorem, given the aforementioned specific choice of the IR subtraction point dictated by perturbation theory to the order of $g^2$, is in fact dominated by the UV asymptotics of the integrand. This is compatible with the universal belief that, nonperturbatively, the IR of the integrand in the rhs of Eqs. \eqref{LETope0} and \eqref{LETope} is exponentially suppressed because of the glueball mass gap \cite{MBL}.\par

\appendix
\section{Nonperturbative UV asymptotics of the OPE} \label{A}

According to the RG, the nonperturbative UV asymptotics of the renormalized OPE coefficients, $C_0$, $C_1$ and $C^{(O_1,O)}_{O}$, for multiplicatively renormalizable operators, $O$ and $O_1$:
\be \label{51}
O(x) O(0) \sim C_0^{(O)}(x)\mathbb{I} + C_1^{(O)}(x) O_1(0) + \cdots
\ee
\be \label{52}
O_1(x) O(0)  \sim C^{(O_1,O)}_{O}(x) O(0) + \cdots
\ee
follows from the associated CS equations in the coordinate representation, which guarantees the absence of contact terms for $x \neq 0$ and, consequently, of additive renormalizations \cite{MBM,MBN,MBG1}:
\be
\label{CSE0}
\left(x\cdot\f{\partial}{\partial x} + \beta(g)\f{\partial}{\partial g} + 2D + 2\gamma_{O}(g)\right)C_0^{(O)}(x) = 0
\ee
\be
\label{CSE1}
\left(x\cdot\f{\partial}{\partial x} + \beta(g)\f{\partial}{\partial g} + 2D - D_1 + 2\gamma_{O}(g) - \gamma_{O_1}(g)\right)C_1^{(O)}(x) = 0
\ee
\be
\label{CSE10}
\left(x\cdot\f{\partial}{\partial x} + \beta(g)\f{\partial}{\partial g} + D_1 + \gamma_{O_1}(g)\right)C^{(O_1,O)}_{O}(x) = 0
\ee
with $D$, $\gamma_O(g)$ and $D_1$, $\gamma_{O_1}(g)$ the canonical and anomalous dimension of the operators $O$ and $O_1$ respectively.\par
The general solutions \cite{MBM,MBN,MBG1} are:
\begin{equation}\label{eqn:pert_general_behavior}
C_0^{(O)}(x) = \frac{1}{x^{2D}}\,\mathcal{G}_0^{(O)}(g(x))\, Z^{(O)2}(x \mu, g(\mu))
\end{equation}
\begin{equation}
C_1^{(O)}(x) = \frac{1}{x^{2D-D_1}}\,\mathcal{G}_1^{(O)}(g(x))\, Z^{(O)2}(x \mu, g(\mu)) Z^{(O_1) -1}(x \mu, g(\mu))
\end{equation}
\begin{equation}
C^{(O_1,O)}_{O}(x) = \frac{1}{x^{D_1}}\,\mathcal{G}^{(O_1,O)}_{O}(g(x))\, Z^{(O_1)}(x \mu, g(\mu))
\end{equation}
which are expressed in terms of the RG-invariant functions, $\mathcal{G}_0^{(O)}$,  $\mathcal{G}_1^{(O)}$ and $\mathcal{G}^{(O_1,O)}_{O}$, of $g(x)$ only, and of the renormalized multiplicative factors, $Z^{(O)}$:
\begin{equation} \label{def}
Z^{(O)}(x \mu, g(\mu)) = \exp \int_{g(\mu)} ^{g(x)}      \frac{ \gamma_O (g) } {\beta(g)} dg 
\end{equation}
determined by the anomalous dimension, $\gamma_O(g)$:
\bea
\gamma_O(g)= - \frac{\partial \log Z^{(O)}}{\partial \log \mu}=-\gamma^{(O)}_{0} g^2 - \gamma^{(O)}_{1} g^4+ \cdots
\eea 
and the beta function, $\beta(g)$: 
\bea
\beta(g)= \frac{\partial g}{\partial \log \mu}=-\beta_{0} g^3 - \beta_1 g^5 +\cdots
\eea 
that can be computed in perturbation theory.\par
The asymptotic expansion of $Z^{(O)}$ for $x \rightarrow 0$ follows from Eq. \eqref{def}:
\bea \label{zeta00}
&Z^{(O)}(x \mu, g(\mu)) \sim \left(\frac{g(x)}{g(\mu)}\right)^{\frac{\gamma_0^{(O)}}{\beta_0}} \exp \left( \frac{\gamma_1^{(O)}  \beta_0 - \gamma_{0}^{(O)} \beta_1}{2 \beta_0^2} (g^2(x)-g^2(\mu))+\cdots \right)
\eea
where the dots represent a series in integer powers of $g^2(x)$ and $g^2(\mu)$ greater than $1$. By an abuse of notation we have set $g(x) \equiv g(x \Lambda_{QCD})$ and  $g(\mu) \equiv g(\mu^{-1}\Lambda_{QCD})$, where within the universal -- i.e., renormalization-scheme independent -- leading and next to leading asymptotic accuracy:
\be \label{alfa}
g^2(x \Lambda_{QCD}) \sim \dfrac{1}{\beta_0 \log(\frac{1}{x^2\Lambda^2_{QCD}})}\left(1-\frac{\beta_1}{\beta_0^2}\frac{\log\log(\frac{1}{x^2\Lambda^2_{QCD}})}{\log(\frac{1}{x^2\Lambda^2_{QCD}})}\right)
\ee 
Indeed, in Eq. \eqref{alfa} we may substitute to $\Lambda_{QCD}$ any finite scale without changing the universal asymptotics.
Thus:
\bea \label{zeta100}
Z^{(O)}(x\mu,g(\mu)) \sim \left(\frac{g(x)}{g(\mu)}\right)^{\frac{\gamma_0^{(O)}}{\beta_0}}  Z^{(O)'}(g(\mu))
\eea
where the constant factor, $Z^{(O)'}(g(\mu))$, is the limit of the exponential in Eq. \eqref{zeta00} as $g(x) \rightarrow 0$. 
For brevity, in writing the universal UV asymptotics of $Z^{(O)}$, we skip systematically the factor of $Z^{(O)'}(g(\mu))$, which is on the order of $1+O(g^2(\mu))$. \par
Hence, the universal UV asymptotics of the OPE coefficients \cite{MBM,MBN,MBG1} is:
\be
\label{CSS0}
C_0^{(O)}(x) \sim \f{1}{x^{2D}} \left(\f{g(x)}{g(\mu)}\right)^{\f{2 \gamma^{(O)}_{0}}{\beta_0}}
\ee
\be
\label{CSS1}
C_1^{(O)}(x) \sim \f{1}{x^{2D-D_1}}g^{2l}(x)\left(\f{g(x)}{g(\mu)}\right)^{\f{2  \gamma^{(O)}_{0}- \gamma^{(O_1)}_{0}}{\beta_0}}
\ee
\be
\label{CSS2}
C^{(O_1,O)}_{O}(x) \sim \f{1}{x^{D_1}}g^{2k}(x)\left(\f{g(x)}{g(\mu)}\right)^{\f{\gamma^{(O_1)}_{0}}{\beta_0}}
\ee
for some integer $l$ and $k$, up to constant normalization factors that can be fixed by perturbation theory.\par
The RG-invariant factors, $g^{2l}(x)$ and $g^{2k}(x)$, which arise from the asymptotics of $\mathcal{G}_1^{(O)}(g(x))$ and $\mathcal{G}^{(O_1,O)}_{O}(g(x))$ respectively, account for the possibility that the 3-point correlator, $\langle O(z) O(0) O_1(x)\rangle$, vanishes to some perturbative order. \par
Indeed, we have shown in Sec. \ref{5} that $\langle O(z) O (0) F^2(x)\rangle$ necessarily starts to the order of $g^2$ in perturbation theory up to contact terms as a consequence of the low-energy theorem, and thus actually vanishes to the lowest order up to contact terms.\par
Instead, the 2-point correlator, $C_0^{(O)}(x)$, of a nontrivial Hermitian operator, $O$, is necessarily nonvanishing \cite{MBM,MBN,MBG1} in the conformal limit of a unitary massless AF QCD-like theory. \par
It follows from Eq. \eqref{F2ren}:
\be
\label{CSS0F2}
C_0^{(S)}(x) \sim \f{N^2-1}{\pi^4 x^8}\left(\f{g(x)}{g(\mu)}\right)^4
\ee
from Eq. \eqref{C1}:
\be \label{CSS1F2}
C_1^{(S)}(x) \sim \f{4 \beta_0}{\pi^2 x^4}g^{2}(x)\left(\f{g(x)}{g(\mu)}\right)^2
\ee
and from Eq. \eqref{d}:
\bea \label{as}
C^{(F^2,O)}_{O}(x) \sim \f{2\gamma_0^{(O)}}{\pi^2x^{4}} g^{2}(x)\left(\f{g(x)}{g(\mu)}\right)^{2}
\eea

\section{Perturbative OPE} \label{B}

\subsection{Perturbative $C_0^{(S)}$ in the momentum representation} \label{B1}

For the reader convenience, we report the relative normalization of the OPE coefficients computed in \cite{Ch, Z1, Z2} for the multiplicatively renormalized
operator, $F^2$, in the $\overline{MS}$ scheme:
\be
F^2(z)F^2(0) \sim 16 C_{0CZ}^{(S)}(z)\mathbb{I} - 4C_{1CZ}^{(S)}(z)F^2(0)
\ee
with respect to our conventions:
\bea
F^2(z)F^2(0) & \sim & C_0^{(S)}(z)\mathbb{I} + C_1^{(S)}(z)F^2(0) 
\eea
Thus:
\bea
C_0^{(S)} & = & 16 \, C_{0CZ}^{(S)} \\
C_1^{(S)} & = & -4 \, C_{1CZ}^{(S)}
\eea
$C_{0CZ}^{(S)}$ has been computed in \cite{Ch,Z1,Z2} to three loops in the momentum representation:
\bea
\label{C0p} 
 C_{0CZ}^{(S)}(p) & = &  \f{N^2-1}{16\pi^2}p^4\left\{ -\frac{\log(\f{p^2}{\mu^2})}{4}+\frac{1}{4}   \right. \nn \\ &{}& \left.
 + \as \left(\frac{11}{48} \ca   \log^2(\f{p^2}{\mu^2}) 
 -\frac{73 \ca \log(\f{p^2}{\mu^2})   }{48}
   -\frac{3 \ca  \zeta_3}{4}+\frac{485 \ca   }{192} \right. \right. \nn \\  &{}& \left.\left.
   -\frac{1}{12} \log^2(\f{p^2}{\mu^2})  \Nf   \tr
   +\frac{7}{12} \log(\f{p^2}{\mu^2})  \Nf \tr
   -\frac{17    \Nf \tr}{16}\right)  \right. \nn \\ &&\left.
  +\as^2 \left(-\frac{121}{576} \ca^2 \log^3(\f{p^2}{\mu^2})   
  +\frac{313}{128} \ca^2 \log^2(\f{p^2}{\mu^2})   
  +\frac{55}{32} \ca^2 \log(\f{p^2}{\mu^2})    \zeta_3  \right. \right. \nn \\  &{}& \left.\left.
  -\frac{37631 \ca^2 \log(\f{p^2}{\mu^2})   }{3456}
   -\frac{2059}{288} \ca^2    \zeta_3  
   +\frac{11}{64} \ca^2  \zeta_4
   +\frac{25}{16}   \ca^2  {\zeta_5}
   +\frac{707201 \ca^2   }{41472} \right. \right. \nn \\  &{}& \left.\left.
   +\frac{11}{72} \ca \log^3(\f{p^2}{\mu^2})    \Nf \tr
   -\frac{85}{48} \ca \log^2(\f{p^2}{\mu^2})    \Nf \tr
   +\frac{1}{8} \ca \log(\f{p^2}{\mu^2})  \Nf   \tr \zeta_3 \right. \right. \nn \\  &{}& \left.\left.
   +\frac{6665}{864} \ca \log(\f{p^2}{\mu^2})    \Nf \tr 
   +\frac{169}{144} \ca  \Nf \tr   \zeta_3
   -\frac{7}{16} \ca  \Nf \tr   \zeta_4 \right. \right. \nn \\  &{}& \left.\left.
   -\frac{7847}{648} \ca  \Nf   \tr
   -\frac{1}{8} {\cf} \log^2(\f{p^2}{\mu^2})  \Nf   \tr
   -\frac{3}{4} {\cf} \log(\f{p^2}{\mu^2})  \Nf \tr   \zeta_3 \right. \right. \nn \\  &{}& \left.\left.
   +\frac{131}{96} {\cf} \log(\f{p^2}{\mu^2})  \Nf   \tr 
   +\frac{41}{24} {\cf}  \Nf \tr   \zeta_3
   +\frac{3}{8} {\cf}  \Nf \tr   \zeta_4 \right. \right. \nn \\  &{}& \left.\left.
   -\frac{5281 {\cf}  \Nf   \tr}{1728}
   -\frac{1}{36} \log^3(\f{p^2}{\mu^2})  \Nf^2   \tr^2
   +\frac{7}{24} \log^2(\f{p^2}{\mu^2})  \Nf^2   \tr^2\right. \right. \nn \\  &{}& \left.\left.
   -\frac{127}{108} \log(\f{p^2}{\mu^2})  \Nf^2   \tr^2 
   +\frac{4715  \Nf^2   \tr^2}{2592}\right) \right\} 
 \eea
with $C_A = N$, $C_F = \f{N^2-1}{2N}$, $T_F = \f{1}{2}$ and $\as = \f{g^2_{YM}(\mu)}{4\pi^2}$. \par
Hence, multiplying Eq. \eqref{C0p} by 16, and expressing $\as$ in terms of the 't Hooft coupling, $g$, we get $C_0^{(S)}$ in the momentum representation:
\bea
\label{C0pFull}
C_{0}^{(S)}(p) & = & \f{N^2-1}{4\pi^2}p^4\left(1 - \log (\f{p^2}{\mu^2}) + g^2(\mu) \left(B_{0,1}^{(S)} - B_{0,2}^{(S)}\log(\f{p^2}{\mu^2}) + \beta_0\log^2(\f{p^2}{\mu^2})\right) \right. \nn \\
&{}& \left. + g^4(\mu)\left(B_{0,3}^{(S)} - B_{0,4}^{(S)}\log(\f{p^2}{\mu^2}) + B_{0,5}^{(S)}\log^2(\f{p^2}{\mu^2}) + \beta_0^2\log^3(\f{p^2}{\mu^2})\right)\right) \nn \\
& + &  \f{N^2-1}{4\pi^2}p^4\left(\log(\f{\Lambda^2}{\mu^2}) + g^2(\mu)\left(B_{0,6}^{(S)}\log(\f{\Lambda^2}{\mu^2}) - \beta_0\log^2(\f{\Lambda^2}{\mu^2})\right) \right. \nn \\
&{}& \left. + g^4(\mu)\left(B_{0,7}^{(S)}\log(\f{\Lambda^2}{\mu^2}) + B_{0,8}^{(S)}\log^2(\f{\Lambda^2}{\mu^2}) + \beta_0^2\log^3(\f{\Lambda^2}{\mu^2})\right)\right)
\eea
with:
\bea
B_{0,1}^{(S)} & = & \f{1}{(4\pi)^2}\left(\f{485}{12} - 12\zeta_3 - \f{17}{2}\f{\Nf}{N}\right) \\
B_{0,2}^{(S)} & = & \f{1}{(4\pi)^2}\left(\f{73}{3} - \f{14}{3}\f{\Nf}{N}\right) \\
B_{0,3}^{(S)} & = & \f{1}{(4\pi)^4}\left(11 \zeta_4+100 \zeta_5-\frac{4118 \zeta_3}{9}+\frac{707201}{648}+\frac{584 \Nf \zeta_3}{9 N} \right. \nn \\
&{}& \left.  -\frac{141395 \Nf}{324 N} +\frac{4715 \Nf^2}{162 N^2} -\frac{6 \Nf \zeta_4}{ N^3}-\frac{82 \Nf \zeta_3}{3 N^3}+\frac{5281 \Nf}{108 N^3}\right) \\
B_{0,4}^{(S)} & = & \f{1}{(4\pi)^4}\left(-110\zeta_3 + \f{37631}{54} - 4\zeta_3\f{\Nf}{N} - \f{6665}{27}\f{\Nf}{N} + 24\zeta_3\f{N^2-1}{2N^2}\f{\Nf}{N} \right. \nn \\ 
&{}& \left. - \f{131}{3}\f{N^2-1}{2N^2}\f{\Nf}{N} + \f{508}{27}\f{\Nf^2}{N^2}\right) \\
B_{0,5}^{(S)} & = & \f{1}{(4\pi)^4}\left(\f{313}{2} - \f{170}{3}\f{\Nf}{N} - 4\f{N^2-1}{2N^2}\f{\Nf}{N} + \f{14}{3}\f{\Nf^2}{N^2}\right) \\
B_{0,6}^{(S)} & = & \f{1}{(4\pi)^2}\left(\f{17}{2} - \f{5}{3}\f{\Nf}{N}\right) \\
B_{0,7}^{(S)} & = & \f{1}{(4\pi)^4}\left(\f{22\zeta_3}{3} + \f{22351}{324} - \f{28\zeta_3}{3}\f{\Nf}{N} - \f{1598}{81}\f{\Nf}{N} + 8\zeta_3\f{N^2-1}{2N^2} \right. \nn \\
&{}& \left. - \f{107}{9}\f{N^2-1}{2N^2}\f{\Nf}{N} + \f{49}{81}\f{\Nf^2}{N^2}\right) \\
B_{0,8}^{(S)} & = & \f{1}{(4\pi)^4}\left(-\f{833}{18} + \f{146}{9}\f{\Nf}{N} + \f{8}{3}\f{N^2-1}{2N^2}\f{\Nf}{N} - \f{10}{9}\f{\Nf^2}{N^2}\right)
\eea
In Eq. \eqref{C0pFull} we have also included the contact terms computed in \cite{Z2}.
 
 \subsection{Perturbative $C_0^{(S)}$ in the coordinate representation} \label{B2}
 
To get $C_0^{(S)}$ in the coordinate representation, we perform the Fourier transform employing the relations \cite{MBM}:
 \begin{eqnarray}
\label{5.28}
&&\int p^4\log\dfrac{p^2}{\mu^2}e^{ip\cdot x}\dfrac{d^{4}p}{(2\pi)^{4}} = -\frac{2^6 3}{\pi ^2 x^8}
\end{eqnarray}
\begin{eqnarray}
\label{5.29}
&&\int p^{4}\log^2 \dfrac{p^{2}}{\mu^{2}}e^{i\cdot x}\dfrac{d^{4}p}{(2\pi)^{4}} = \frac{2^7 3}{\pi ^2 x^8} \left(\log (\mu^2 x^2)-\frac{10}{3}+2 \gamma_E -\log 4 \right)
\end{eqnarray}
\bea
\label{5.30}
 \int p^{4}\log^3 \dfrac{p^{2}}{\mu^{2}} e^{ip\cdot x} \dfrac{d^{4}p}{(2\pi)^{4}}  & = & \frac{2^6 3}{\pi ^2 x^8} \left(-3 \left(\log 4-\log (\mu^2 x^2)\right)^2 \right. \nonumber \\
&& \left. +(20-12 \gamma_E ) \log( \mu^2 x^2)   -12 \gamma_E ^2 \right. \nn \\
&& \left. -\frac{51}{2} +40 \gamma_E -(20-12 \gamma_E ) \log 4 \right)
\eea
It follows $C_0^{(S)}$ in the coordinate representation:
\bea
\label{C0x1}
C_0^{(S)}(x) & = &  \f{N^2-1}{x^8}\f{48}{\pi^4} \left(1 + g^2(\mu) (A_{0,1}^{(S)} + 2\beta_0 \log(x^2\mu^2))  \right. \nn \\
&{}&\left. + g^4(\mu) (A_{0,2}^{(S)} + A_{0,3}^{(S)}\log(x^2\mu^2) + 3\beta_0^2 \log^2(x^2\mu^2)) \right) \nn \\
& + & \Delta^2 \delta^{(4)}(x)\dfrac{N^2-1}{4\pi^2}\left(1 + \log(\f{\Lambda^2}{\mu^2}) + g^2(\mu)\left(A_{0,4}^{(S)} + A_{0,5}^{(S)}\log(\f{\Lambda^2}{\mu^2})  \right. \right. \nn \\
&{}& - \left.\left.\beta_0 \log^2(\f{\Lambda^2}{\mu^2})\right) + g^4(\mu)\left(A_{0,6}^{(S)} + A_{0,7}^{(S)}\log(\f{\Lambda^2}{\mu^2}) + A_{0,8}^{(S)}\log^2(\f{\Lambda^2}{\mu^2}) \right. \right. \nn \\
&{}& + \left. \left. \beta_0^2 \log^3(\f{\Lambda^2}{\mu^2 })\right)\right) 
\eea
with $A_{0,1}^{(S)}$, $A_{0,2}^{(S)}$, $A_{0,3}^{(S)}$ scheme dependent:
\bea
A_{0,1}^{(S)} & = & \f{1}{(4\pi)^2}\left(\frac{132 \gamma_E -1-132 \log2}{9}-\frac{2\Nf(12 \gamma_E +1-12 \log2)}{9 N}\right) \\
A_{0,2}^{(S)} & = & \f{1}{(4\pi)^4}\left(\frac{-98+4356 \log^2 2-8712 \gamma_E  \log2-2382 \log2}{27} \right. \nn \\
&& \left.+ \f{-2970 \zeta_3 +4356 \gamma_E ^2+2382 \gamma_E}{27} \right. \nn \\
&& \left. -\f{\Nf \left(-2112 \log2-6336 \gamma_E  \log2 \right)}{54 N} \right.\nn \\
&&\left.-\frac{\Nf \left(-432 \zeta_3 +3168 \gamma_E ^2+2112 \gamma_E +121+3168 \log^2 2\right)}{54 N}\right.\nn \\
&&\left.+\frac{2\Nf^2 \left(72 \gamma_E ^2+12 \gamma_E -13+72 \log^2 2 -144 \gamma_E  \log2 -12 \log2\right)}{27 N^2}\right. \nn \\
&&\left. +\frac{\Nf (-24 \zeta_3+16 \gamma_E +17-16 \log2)}{2 N^3}\right)  \\
A_{0,3}^{(S)} & = & \f{1}{(4\pi)^4}\left(\frac{1452 \gamma_E +397-1452 \log2}{9}   -\frac{88 \Nf (6 \gamma_E +2-6 \log2)}{9 N}\right. \nn \\
&&\left. +\frac{4\Nf^2 (12 \gamma_E +1-12 \log2)}{9  N^2} + \frac{4\Nf}{  N^3}\right)
\eea
and $A_{0,4}^{(S)}$, $A_{0,5}^{(S)}$, $A_{0,6}^{(S)}$, $A_{0,7}^{(S)}$, $A_{0,8}^{(S)}$ coefficients of the contact terms:
\bea
\label{C0CT}
A_{0,4}^{(S)} & = & \f{1}{(4\pi)^2}\left(\f{485}{12}-12\zeta_3-\f{17\Nf}{2N}\right)  \\
A_{0,5}^{(S)} & = & \f{1}{(4\pi)^2}\left(\f{17}{2}-\f{5}{3}\f{\Nf}{N}\right)  \\
A_{0,6}^{(S)} & = & \f{1}{(4\pi)^4}\left(11 \zeta_4+100 \zeta_5-\frac{4118 \zeta_3}{9}+\frac{707201}{648}+\frac{584 \Nf \zeta_3}{9 N} \right. \nn \\
&{}& \left.  -\frac{141395 \Nf}{324 N} +\frac{4715 \Nf^2}{162 N^2} -\frac{6 \Nf \zeta_4}{ N^3}-\frac{82 \Nf \zeta_3}{3 N^3}+\frac{5281 \Nf}{108 N^3}\right)  \\
A_{0,7}^{(S)} & = &\f{1}{(4\pi)^4}\left(\f{22\zeta_3}{3} + \f{22351}{324} - \f{28\zeta_3}{3}\f{\Nf}{N} - \f{1598}{81}\f{\Nf}{N} + 8\zeta_3\f{N^2-1}{2N^2} \right. \nn \\
&{}& \left. - \f{107}{9}\f{N^2-1}{2N^2}\f{\Nf}{N} + \f{49}{81}\f{\Nf^2}{N^2}\right) \\
A_{0,8}^{(S)} & = & \f{1}{(4\pi)^4}\left(-\f{833}{18} + \f{146}{9}\f{\Nf}{N} + \f{8}{3}\f{N^2-1}{2N^2}\f{\Nf}{N} - \f{10}{9}\f{\Nf^2}{N^2}\right)
\eea

\subsection{Verifying the UV asymptotics of $C_0^{(S)}$ by a change of renormalization scheme in perturbation theory} \label{B3}

We skip the contact terms, and we multiply Eq. \eqref{C0x1} by $g^4$ that is equivalent to consider the OPE of $g^2 F^2$.
Then, following \cite{MBM} we change the renormalization scheme redefining the coupling constant:
\be
g^2_{ab}(\mu) = g^2(\mu)(1 + a g^2(\mu) + b g^4(\mu))
\ee
with $a$ and $b$ such that the constant finite parts \footnote{We define the divergent parts as the terms that, after setting $\mu=\Lambda$, become divergent as $\Lambda \rightarrow \infty$. The constant finite parts are the remaining constant terms.} of $g^4 C^{(S)}_0$ vanish to the order of $g^8$. Eq. \eqref{C0x1} becomes:
\begin{small}
\bea
\label{C0x2}
g^4_{ab}(\mu)C_0^{(S)}(x) & = & \f{N^2-1}{\pi^4}\f{48 g_{ab}^4(\mu)}{ x^8}\left(1 + g_{ab}^2(\mu)(A_{0,1}^{(S)} -2a  +  2\beta_0 \log(x^2\mu^2))  \right.  \nn \\
&& +\left.
 g_{ab}^4(\mu)(A_{0,2}^{(S)} + 5a^2 - 2b -3a A_{0,1}^{(S)}  \right. \nn \\
&& \left.
+ (A_{0,3}^{(S)}+ \frac{a (2 \Nf-11 N)}{8 \pi ^2 N}) \log(x^2\mu^2) + 3\beta_0^2 \log^2(x^2\mu^2))\right) 
\eea
\end{small}
with:
\be
a = \f{A_{0,1}^{(S)}}{2}
\ee
\be
b = \f{4 A_{0,2}^{(S)} - A_{0,1}^{(S)2}}{8}
\ee
Remarkably, the coefficient of the term $g_{ab}^4(\mu) \log(x^2\mu^2)$ is \cite{MBM} now:
\bea
&A_{0,3}^{(S)} & + \frac{a (2 \Nf-11 N)}{8 \pi ^2 N} \nn \\ 
& = & A_{0,3}^{(S)} + \frac{A_{0,1}^{(S)} (2 \Nf-11 N)}{16 \pi ^2 N} \nn \\
& = & \f{1}{(4\pi)^4}\left(\frac{1452 \gamma_E +397-1452 \log2}{9}   -\frac{88 \Nf (6 \gamma_E +2-6 \log2)}{9 N}\right. \nn \\
&&\left. +\frac{4\Nf^2 (12 \gamma_E +1-12 \log2)}{9  N^2} + \frac{4\Nf}{  N^3}\right) \nn \\
&&  +\frac{(2 \Nf-11 N)}{16 \pi ^2 N}\f{1}{(4\pi)^2}\left(\frac{132 \gamma_E -1-132 \log2}{9}-\frac{2\Nf(12 \gamma_E +1-12 \log2)}{9 N}\right) \nn \\
& = & \frac{34 N^3-13 N^2 \Nf+3 \Nf}{192 \pi ^4 N^3} = 4\beta_1
\eea
It follows:
\bea
\label{C0x3}
g^4_{ab}(\mu)C_0^{(S)}(x) & = & \f{N^2-1}{\pi^4}\f{48 g_{ab}^4(\mu)}{x^8}\left(1 + g_{ab}^2(\mu) 2\beta_0 \log(x^2\mu^2) \right. \nn \\
&& \left.
+ g^4_{ab}(\mu)( 4\beta_1\log(x^2\mu^2)  + 3\beta_0^2 \log^2(x^2\mu^2))\right)
\eea
Thus, manifestly $\gamma^{(F^2)}_1= 4 \beta_1$ in this scheme, according to Eq. \eqref{ad}.\par
In order to verify Eq. \eqref{CSS0F2} we should express Eq. \eqref{C0x3} in terms of $g(x)$, which reads to two loops in $\overline{MS}$-like schemes:
\bea
\label{gr}
g^2(x) & = & g^2(\mu)\left(1 + g^2(\mu)\beta_0\log(x^2\mu^2) + g^4(\mu)(\beta_1\log(x^2\mu^2) \right. \nn \\
&& \left.  + \beta_0^2\log^2(x^2\mu^2))\right)
\eea
Hence, to two loops:
\bea
\label{grun}
g^4(x) & = & g^4(\mu)\left(1 + g^2(\mu) 2\beta_0\log(x^2\mu^2) + g^4(\mu)(2\beta_1\log(x^2\mu^2) \right. \nn \\
&& \left.  + 3\beta_0^2\log^2(x^2\mu^2))\right)
\eea
This cannot be done immediately, because the coefficient of the term $g_{ab}^4(\mu)\log(x^2\mu^2)$ in Eq. \eqref{C0x3} is $4\beta_1$
instead of $2\beta_1$ in Eq. \eqref{grun}. Following \cite{MBM} we insert the identity:
\be
1 = \f{1+2\frac{\beta_1}{\beta_0}g^2_{ab}(\mu)}{1+2\frac{\beta_1}{\beta_0}g^2_{ab}(x)}\f{1+2\frac{\beta_1}{\beta_0}g^2_{ab}(x)}{1+2\frac{\beta_1}{\beta_0}g^2_{ab}(\mu)}
\ee
in Eq. \eqref{C0x3} to the relevant order:
\bea
\label{C0x4}
g^4_{ab}(\mu)C_0^{(S)}(x) & = & \f{1+2\frac{\beta_1}{\beta_0}g^2_{ab}(\mu)}{1+2\frac{\beta_1}{\beta_0}g^2_{ab}(x)}\f{1+2\frac{\beta_1}{\beta_0}g^2_{ab}(x)}{1+2\frac{\beta_1}{\beta_0}g^2_{ab}(\mu)}\f{N^2-1}{\pi^4} \f{48 g_{ab}^4(\mu)}{x^8}\nn \\
&& \left(1 + g_{ab}^2(\mu) 2\beta_0 \log(x^2\mu^2) + g^4_{ab}(\mu)( 4\beta_1\log(x^2\mu^2) + 3\beta_0^2 \log^2(x^2\mu^2))\right)  \nn \\
& = & \f{1+2\frac{\beta_1}{\beta_0}g^2_{ab}(x)}{1+2\frac{\beta_1}{\beta_0}g^2_{ab}(\mu)}\f{N^2-1}{\pi^4}\f{48 g_{ab}^4(\mu)}{x^8}\left(1 + g_{ab}^2(\mu) 2\beta_0 \log(x^2\mu^2) \right. \nn \\
&& \left. + g^4_{ab}(\mu)( 2\beta_1\log(x^2\mu^2) + 3\beta_0^2 \log^2(x^2\mu^2))\right) \nn \\
& = & \f{N^2-1}{\pi^4}\f{48}{x^8}g^4_{ab}(x)\left(1+ 2\f{\beta_1}{\beta_0}g^2_{ab}(x) - 2\f{\beta_1}{\beta_0}g^2_{ab}(\mu)\right) 
\eea
where we have expanded to the order of $g^4$:
\be
\f{1+2\frac{\beta_1}{\beta_0}g^2_{ab}(\mu)}{1+2\frac{\beta_1}{\beta_0}g^2_{ab}(x)} = 1 - 2 \beta_1 g^4_{ab}(\mu)\log(x^2\mu^2) 
\ee
and to the order of $g^2$:
\be
\f{1+2\frac{\beta_1}{\beta_0}g^2_{ab}(x)}{1+2\frac{\beta_1}{\beta_0}g^2_{ab}(\mu)} = 1 + 2\f{\beta_1}{\beta_0}g^2_{ab}(x)-2\f{\beta_1}{\beta_0}g^2_{ab}(\mu)
\ee
The factor of $\left(1+ 2\f{\beta_1}{\beta_0}g^2_{ab}(x) - 2\f{\beta_1}{\beta_0}g^2_{ab}(\mu)\right) $ in the last line of Eq. \eqref{C0x4} arises according to Eq. \eqref{zeta00}, as it is verified by setting $\gamma^{(F^2)} _{0}=2 \beta_0$ and $\gamma^{(F^2)} _1=4 \beta_1$ in Eq. \eqref{zeta00}:
\bea \label{P}
&& \langle \frac{g^2_{ab}(\mu)}{2} F^2(x)  \frac{g^2_{ab}(\mu)}{2}  F^2(0) \rangle \sim   \frac{g^4_{ab}(\mu)}{4} C_0^{(S)}(x) \nonumber \\
&& \sim  \f{N^2-1}{\pi^4}\f{12}{x^8} g^4 _{ab}(\mu) Z^{(F^2)2}(x \mu, g_{ab}(\mu)) \nonumber \\
&& \sim  \f{N^2-1}{\pi^4}\f{12}{x^8}  g^4 _{ab}(\mu) \left(\frac{g_{ab}(x)}{g _{ab}(\mu)}\right)^{\frac{2\gamma_0^{(F^2)}}{\beta_0}} \exp \left( \frac{\gamma_1^{(F^2)}  \beta_0 - \gamma_{0}^{(F^2)} \beta_1}{\beta_0^2} (g^2_{ab}(x)-g^2_{ab}(\mu))+\cdots \right) \nonumber \\
&& \sim  \f{N^2-1}{\pi^4}\f{12}{x^8}g^4_{ab}(x)\left(1+ 2\f{\beta_1}{\beta_0}g^2_{ab}(x) - 2\f{\beta_1}{\beta_0}g^2_{ab}(\mu) + \cdots \right) 
\eea
Therefore, Eq. \eqref{C0x4} agrees with Eq. \eqref{CSS0F2}. \par
The computation in Eq. \eqref{P} exhibits the scale dependence in massless QCD of the 2-point correlator of $ \frac{g^2}{2} F^2$ in the scheme where manifesly $\gamma^{(F^2)} _1=4 \beta_1$. Hence, the aforementioned operator is not RG invariant. \par 
As an aside, $\frac{g^2}{2N} F^2$ is the YM Lagrangian density with the canonical normalization (Subsec. \ref{2.3}). \par
We may remove the scale-dependent term, $ - 2\f{\beta_1}{\beta_0}g^2_{ab}(\mu)$, in the last line of Eqs. \eqref{C0x4} and \eqref{P} multiplying them by $\beta_0^2(1 + \f{\beta_1}{\beta_0}g^2_{ab}(\mu))^2$ that makes  $g^4_{ab}(\mu)C_0^{(S)}$ RG invariant to the relevant perturbative order \cite{MBM}. Hence, we get:
\bea
\left(\f{\beta(g)}{g}\right)^2 C_0^{(S)}(x) & \sim & \f{N^2-1}{\pi^4}\f{48\beta_0^2}{x^8} g^4_{ab}(x)\left(1 + 2\f{\beta_1}{\beta_0}g^2_{ab}(x)\right) \nn \\
& \sim & \f{N^2-1}{\pi^4}\f{48\beta_0^2}{x^8} g^4_{ab}(x)
\eea
according to the nonperturbative asymptotics in Eq. \eqref{F2UV}.

\subsection{Perturbative $C_1^{(S)}$ in the momentum representation} \label{B4}

$C_{1CZ}^{(S)}$ has been computed in \cite{Ch,Z1,Z2} to three loops in the momentum representation in the $\overline{MS}$ scheme:
\begin{small}
\bea
\label{C1GG_3l_full}
   C_{1CZ}^{(S)}(p) =
&-&1
+\as \left\{
-\frac{49 \ca}{36}+\frac{5 \Nf \tr}{9}
-\log(\f{p^2}{\mu^2}) \left(\frac{\Nf \tr}{3}-\frac{11   \ca}{12}\right)
\right\}\nonumber\\
&+&\as^2 \left\{
   \frac{33 \ca^2 \zeta_3}{8}
   -\frac{11509 \ca^2}{1296}
   +\frac{3}{2} \ca \Nf \tr \zeta_3
   +\frac{3095 \ca \Nf   \tr}{648}
   -3 \cf \Nf \tr \zeta_3 \right.\nonumber\\&{}&\left.
   +\frac{13 \cf \Nf   \tr}{4}
   -\frac{25 \Nf^2 \tr^2}{81}
   -\log(\f{p^2}{\mu^2})   \left(
-\frac{1151 \ca^2}{216}
+\frac{97 \ca \Nf \tr}{27}      
+\cf \Nf \tr              \right.\right.\nonumber\\&{}&\left.\left.
-\frac{10 \Nf^2   \tr^2}{27}
\right)
   +\log^2(\f{p^2}{\mu^2}) \left(
   -\frac{121 \ca^2}{144}
   +\frac{11 \ca \Nf   \tr}{18}
   -\frac{\Nf^2 \tr^2}{9}
   \right)\right.\nonumber\\&{}&\left.
   +\log(\f{\Lambda^2}{\mu^2})\left[
      -\frac{17 \ca^2}{24 }
      +\frac{5 \ca   \Nf \tr}{12 }
      +\frac{\cf \Nf \tr}{4 }\right]
   \right\}  \nn \\ 
&+&\as^3 \left\{
\frac{5315 \ca^3   \zeta_3}{144}
-\frac{55 \ca^3 \zeta_5}{8}
-\frac{9775633 \ca^3}{186624}
-\frac{263}{144} \ca^2 \Nf \tr \zeta_3      \right.\nonumber\\&{}&\left.
-5   \ca^2 \Nf \tr \zeta_5                 
+\frac{1299295 \ca^2 \Nf \tr}{31104}
-\frac{331}{16} \ca \cf \Nf \tr \zeta_3
-\frac{15}{2} \ca   \cf \Nf \tr \zeta_5     \right.\nonumber\\&{}&\left.
+\frac{35707 \ca \cf \Nf \tr}{1152}            
-\frac{121}{36} \ca \Nf^2 \tr^2   \zeta_3 
-\frac{116773 \ca \Nf^2 \tr^2}{15552}
-9 \cf^2 \Nf   \tr \zeta_3     \right.\nonumber\\&{}&\left.
+15 \cf^2 \Nf \tr \zeta_5
-\frac{45}{16} \cf^2 \Nf \tr                
+\frac{13}{2} \cf \Nf^2 \tr^2   \zeta_3 
-\frac{2399}{288} \cf \Nf^2 \tr^2
+\frac{125 \Nf^3 \tr^3}{729}                  \right.\nonumber\\&{}&\left.
-\log(\f{p^2}{\mu^2}) \left(\frac{363 \ca^3   \zeta_3}{32}
-\frac{360325 \ca^3}{10368}
+\frac{55757 \ca^2 \Nf \tr}{1728}
-\frac{33}{4} \ca \cf \Nf   \tr \zeta_3              \right.\right.\nonumber\\&{}&\left.\left.
+\frac{2527}{192} \ca \cf \Nf \tr 
-\frac{3}{2} \ca \Nf^2 \tr^2   \zeta_3
-\frac{2057}{288} \ca \Nf^2 \tr^2
-\frac{9}{32} \cf^2 \Nf \tr              \right.\right.\nonumber\\&{}&\left.\left.
+3 \cf \Nf^2 \tr^2   \zeta_3
-\frac{209}{48} \cf \Nf^2 \tr^2
+\frac{25 \Nf^3 \tr^3}{81}\right)            
+\log^2(\f{p^2}{\mu^2}) \left(
-\frac{1793   \ca^3}{216}          \right.\right.\nonumber\\&{}&\left.\left.
+\frac{273}{32} \ca^2 \Nf \tr
+\frac{55}{32} \ca \cf \Nf \tr   
-\frac{181}{72}   \ca \Nf^2 \tr^2            
-\frac{5}{8} \cf \Nf^2 \tr^2
+\frac{5 \Nf^3 \tr^3}{27}\right)                 \right.\nonumber\\&{}&\left.
-\log^3(\f{p^2}{\mu^2})   \left(
-\frac{1331 \ca^3}{1728}
+\frac{121}{144} \ca^2 \Nf \tr
-\frac{11}{36} \ca \Nf^2   \tr^2
+\frac{\Nf^3 \tr^3}{27}\right)                 \right.\nonumber\\&{}&\left.
+\log(\f{\Lambda^2}{\mu^2})\left[
\frac{1415 \ca^2 \Nf \tr}{864 }
-\frac{2857 \ca^3}{1728 }
+\frac{205 \ca   \cf \Nf \tr}{288 }
-\frac{79 \ca \Nf^2 \tr^2}{432 }           \right.\right.\nonumber\\&{}&\left.\left.
 -\frac{\cf^2 \Nf \tr}{16 }-\frac{11 \cf \Nf^2 \tr^2}{72 }   \right] + 
\log^2(\f{\Lambda^2}{\mu^2})\left[
-\frac{89 \ca^2 \Nf \tr}{144   }
+\frac{187 \ca^3}{288 }   \right.\right.\nonumber\\&{}& \left.\left.
-\frac{11 \ca \cf \Nf \tr}{48 }
+\frac{5 \ca \Nf^2   \tr^2}{36 } +\frac{\cf \Nf^2   \tr^2}{12 } \right] \right\}
\eea
\end{small}
It follows $C_1^{(S)}$ in the momentum representation:
\bea
\label{C1Full}
C_1^{(S)}(p) & = & 4 + g^2(\mu)\left(B_{1,1}^{(S)} + \beta_0\log(\f{p^2}{\mu^2})\right) + g^4(\mu)\left(B_{1,2}^{(S)} + B_{1,3}^{(S)}\log(\f{p^2}{\mu^2})  \right. \nn \\
&{}& \left.  -\beta_0^2\log^2(\f{p^2}{\mu^2}) + 4\beta_1\log(\f{\Lambda^2}{\mu^2})\right) + g^6(\mu)\left(B_{1,4}^{(S)} + B_{1,5}^{(S)}\log(\f{p^2}{\mu^2})  \right. \nn \\
&{}& \left. + B_{1,6}^{(S)}\log^2(\f{p^2}{\mu^2}) -\beta_0^3\log^3(\f{p^2}{\mu^2}) + 8\beta_2\log(\f{\Lambda^2}{\mu^2}) - 4\beta_0\beta_1\log^2(\f{\Lambda^2}{\mu^2})\right) \nn \\
\eea
with:
\bea
B_{1,1}^{(S)} & = & \f{1}{(4\pi)^2}\left(\f{196}{9} - \f{40}{9}\f{\Nf}{N}\right) \\
B_{1,2}^{(S)} & = & \f{1}{(4\pi)^4}\left(-\f{264\zeta_3}{2} + \f{46036}{81} - \f{12380}{81}\f{\Nf}{N} - 48\zeta_3\f{\Nf}{N} + \f{8}{3}\zeta_3\f{N^2-1}{2N^2}\f{\Nf}{N} \right. \nn \\
&{}& \left. - 104\f{N^2-1}{2N^2}\f{\Nf}{N} + \f{400}{81}\f{\Nf^2}{N^2}\right) \\
B_{1,3}^{(S)} & = & \f{1}{(4\pi)^4}\left(-\f{9208}{27} + \f{3104}{27}\f{\Nf}{N} + 16\f{N^2-1}{N^2}\f{\Nf}{N} - \f{160}{27}\f{\Nf^2}{N^2}\right) \\
B_{1,4}^{(S)} & = & \f{1}{(4\pi)^6}\left( \frac{85040 \zeta_3}{9}- 1760\zeta_5-\frac{9775633}{729} -\frac{16612 \Nf \zeta_3}{9 N} -\frac{640 \Nf \zeta_5}{N}\right. \nn \\
&{}& \left. +\frac{3518939 \Nf}{486 N}-\frac{64\Nf^2 \zeta_3}{9 N^2}-\frac{181546 \Nf^2}{243 N^2} +\frac{1900 \Nf \zeta_3}{N^3}-\frac{480 \Nf \zeta_5}{N^3}-\frac{32467 \Nf}{18 N^3}\right. \nn \\
&{}& \left. + \frac{4000 \Nf^3}{729 N^3} -\frac{52 \Nf^2 \zeta_3}{N^4}+\frac{2399 \Nf^2}{9 N^4} -\frac{288 \Nf \zeta_3}{N^5}+\frac{480 \Nf \zeta_5}{ N^5}-\frac{90 \Nf}{N^5}\right) \\
B_{1,5}^{(S)} & = & \f{1}{(4\pi)^6}\left(-2904 \zeta_3+\frac{720650}{81} +\frac{528 \Nf \zeta_3}{N}-\frac{134014 \Nf}{27 N} +\frac{5368 \Nf^2}{9 N^2} \right. \nn \\
&{}& \left. -\frac{800 \Nf^3}{81 N^3}-\frac{528 \Nf \zeta_3}{ N^3}+\frac{2473 \Nf}{3 N^3} +\frac{96 \Nf^2 \zeta_3}{N^4}-\frac{418 \Nf^2}{3 N^4} + \frac{9 \Nf}{N^5}\right) \\
B_{1,6}^{(S)} & = & \f{1}{(4\pi)^6}\left(\frac{57376}{27}-\frac{1202 \Nf}{N} +\frac{1628 \Nf^2}{9 N^2} -\frac{160 \Nf^3}{27 N^3}+\frac{110 \Nf}{N^3}-\frac{20 \Nf^2}{N^4}\right)
\eea
As demonstrated in \cite{Z2}, the divergent contact terms in Eq. \eqref{C1Full} are expressed in terms of the coefficients of the QCD beta function:
\bea
\beta_0 & = & \f{1}{(4\pi)^2}\left(\f{11}{3} - \f{2}{3}\f{\Nf}{N}\right) \\
\beta_1 & = & \f{1}{(4\pi)^4}\left(\f{34}{3} - \f{13}{3}\f{\Nf}{N} +\f{\Nf}{N^3}\right) \\
\beta_2 & = & \f{1}{(4\pi)^6}\left( \f{2857}{54} - \f{1709}{54}\f{\Nf}{N} + \f{56}{27}\f{\Nf^2}{N^2} +\f{187}{36}\f{\Nf}{N^3} - \f{11}{18}\f{\Nf^2}{N^4} + \f{\Nf}{4N^5}\right)
\eea

 \subsection{Perturbative $C_1^{(S)}$ in the coordinate representation} \label{B5}

The Fourier transform of Eq. \eqref{C1GG_3l_full} is:
\begin{small}
\bea
\label{C1x1}
C_1^{(S)}(x) & = & \f{4\beta_0}{\pi^2x^4}g^2(\mu)\left(1 + g^2(\mu)(A_{1,1}^{(S)} + 2\beta_0\log(x^2\mu^2)) + g^4(\mu)(A_{1,2}^{(S)}   \right. \nn \\
&{}& \left. + A_{1,3}^{(S)}\log(x^2\mu^2) + 3\beta_0^2\log^2(x^2\mu^2))\right) + \delta^{(4)}(x)\left(4 + g^2(\mu)A_{1,4}^{(S)}  \right. \nn \\
&{}& \left. + g^4(\mu)\left(A_{1,5}^{(S)} + 4\beta_1\log(\f{\Lambda^2}{\mu^2}) \right) + g^6(\mu)\left(A_{1,6}^{(S)} + 8\beta_2\log(\f{\Lambda^2}{\mu^2}) \right. \right. \nn \\
&{}& \left. \left. - 4\beta_0\beta_1\log^2(\f{\Lambda^2}{\mu^2})\right)\right)
\eea
\end{small}
with $A_{1,1}^{(S)}$, $A_{1,2}^{(S)}$, $A_{1,3}^{(S)}$ scheme dependent:
\bea
A_{1,1}^{(S)} & = & \f{4}{\beta_0}\f{1}{(4\pi)^4}\left(\frac{363 \gamma_E +394-363 \log 2}{27}-\frac{\Nf (132 \gamma_E +155-132 \log 2)}{27 N} \right. \nn \\
&& \left. +\frac{4\Nf^2 (3 \gamma_E +1-3\log 2)}{27 N^2}+\frac{\Nf}{N^3}\right)  \\
A_{1,2}^{(S)} & = & \f{16}{\beta_0}\f{1}{(4\pi)^6}\left(\frac{-117612 \zeta_3+95832 \gamma_E ^2+248424 \gamma_E +188197+95832 \log ^2 2}{2592} \right. \nn \\
&& \left. - \f{191664 \gamma_E  \log 2+248424 \log 2}{2592}  -\frac{\Nf \left(-7128 \zeta_3+17424 \gamma_E ^2+47484 \gamma_E\right)}{864 N} \right. \nn \\
&& \left. + \f{\Nf\left(34553+17424 \log ^2 2-34848 \gamma_E  \log 2-47484 \log 2\right)}{864 N} \right. \nn \\
&& \left. +\frac{11 \Nf^2 (4 \gamma_E +3-4 \log 2) (3 \gamma_E +4-3\log 2)}{36 N^2} -\frac{1152 \gamma_E ^2 \Nf^3}{5184 N^3} \right. \nn \\
&&\left. + \f{\Nf\left(+768 \gamma_E  \Nf^2-160 \Nf^2-768 \Nf^2 \log 2+1152 \Nf^2 \log^2 2\right)}{5184 N^3} \right. \nn \\ 
&& \left. + \f{\Nf\left(-2304 \gamma_E  \Nf^2 \log 2+42768 \zeta_3-35640 \gamma_E -48951+35640 \log 2\right)}{5184 N^3} \right. \nn \\
&& \left. -\frac{\Nf^2 (-144 \zeta_3+120 \gamma_E +149-120 \log 2)}{96 N^4} + \frac{9 \Nf}{64 N^5}\right) \\
A_{1,3}^{(S)} & = & \f{16}{\beta_0}\f{1}{(4\pi)^6}\left(\frac{11 (726 \gamma_E +941-726 \log 2)}{216}-\frac{\Nf (968 \gamma_E +1319-968 \log 2)}{48 N} \right. \nn \\
&&\left. +\frac{11 \Nf^2 (72 \gamma_E +75-48 \log (2)-24 \log 2)}{216 N^2}\right. \nn \\
&& \left. +\frac{-96 \gamma_E  \Nf^3-32 \Nf^3+96 \Nf^3 \log 2+1485 \Nf}{432 N^3} -\frac{5 \Nf^2}{8 N^4}\right)
\eea
and $A_{1,4}^{(S)}$, $A_{1,5}^{(S)}$, $A_{1,6}^{(S)}$ coefficients of the contact terms:
\bea
A_{1,4}^{(S)} & = & \f{1}{(4\pi)^2}\left(\f{196}{9} - \f{40}{9}\f{\Nf}{N}\right) \\
A_{1,5}^{(S)} & = & \f{1}{(4\pi)^4}\left(-\f{264\zeta_3}{2} + \f{46036}{81} - \f{12380}{81}\f{\Nf}{N} - 48\zeta_3\f{\Nf}{N} + \f{8}{3}\zeta_3\f{N^2-1}{2N^2}\f{\Nf}{N} \right. \nn \\
&{}& \left. - 104\f{N^2-1}{2N^2}\f{\Nf}{N} + \f{400}{81}\f{\Nf^2}{N^2}\right)  \\
 A_{1,6}^{(S)} & = & \f{1}{(4\pi)^6}\left( \frac{85040 \zeta_3}{9}- 1760\zeta_5-\frac{9775633}{729} -\frac{16612 \Nf \zeta_3}{9 N} -\frac{640 \Nf \zeta_5}{N}\right. \nn \\
&{}& \left. +\frac{3518939 \Nf}{486 N}-\frac{64\Nf^2 \zeta_3}{9 N^2}-\frac{181546 \Nf^2}{243 N^2} +\frac{1900 \Nf \zeta_3}{N^3}-\frac{480 \Nf \zeta_5}{N^3}-\frac{32467 \Nf}{18 N^3}\right. \nn \\
&{}& \left. + \frac{4000 \Nf^3}{729 N^3} -\frac{52 \Nf^2 \zeta_3}{N^4}+\frac{2399 \Nf^2}{9 N^4} -\frac{288 \Nf \zeta_3}{N^5}+\frac{480 \Nf \zeta_5}{ N^5}-\frac{90 \Nf}{N^5}\right)
\eea

\subsection{Verifying the UV asymptotics of $C_1^{(S)}$ by a change of renormalization scheme in perturbation theory} \label{B6}

As for $C_0^{(S)}$, we skip the contact terms, and we multiply Eq. \eqref{C1x1} by $g^2$ that is equivalent to consider the OPE of $g^2F^2$. Following \cite{MBM} we change the renormalization scheme redefining the coupling constant:
\be
g^2_{cd}(\mu) = g^2(\mu)(1 + cg^2(\mu) + dg^4(\mu))
\ee
Then, Eq. \eqref{C1x1} becomes:
\bea \label{new}
g^2_{cd}(\mu)C_1^{(S)}(x) & = & \f{4\beta_0}{\pi^2 x^4}g^4_{cd}(\mu)\left(1 + g^2_{cd}(\mu) (A_{1,1}^{(S)} - 2c + 2\beta_0\log(x^2\mu^2)) \right. \nn \\
&{}&\left. + g^4_{cd}(\mu) (5c^2-2d-3cA_{1,1}^{(S)} + A_{1,2}^{(S)} \right. \nn \\
&{}& \left.
+ (A_{1,3}^{(S)} -6c\beta_0 )\log(x^2\mu^2)  +3\beta_0\log^2(x^2\mu^2) \right) 
\eea
We fix $c$ and $d$ requiring that the finite parts of $g^2 C_1^{(S)}$ vanish to the order of $g^8$ in the new renormalization scheme:
\be
c = \f{A_{1,1}^{(S)}}{2}
\ee
\be
d = \f{4A_{1,2}^{(S)} - A_{1,1}^{(S)2}}{8}
\ee
Remarkably, as for $g_{ab}^4 C_0^{(S)}$, in this renormalization scheme the coefficient of the term $g_{cd}^4(\mu) \log(x^2\mu^2)$ in Eq. \eqref{new} is proportional to the second coefficient of the QCD beta function:
\bea
& A_{1,3}^{(S)} & - 6c\beta_0 \nn \\
& = & A_{1,3}^{(S)} - 3A_{1,1}^{(S)}\beta_0 \nn \\
& = &\f{16}{\beta_0}\f{1}{(4\pi)^6}\left(\frac{11 (726 \gamma_E +941-726 \log 2)}{216}-\frac{\Nf (968 \gamma_E +1319-968 \log 2)}{48 N} \right. \nn \\
&&\left. +\frac{11 \Nf^2 (72 \gamma_E +75-48 \log 2 -24 \log 2)}{216 N^2}\right. \nn \\
&& \left. +\frac{-96 \gamma_E  \Nf^3-32 \Nf^3+96 \Nf^3 \log 2+1485 \Nf}{432 N^3} -\frac{5 \Nf^2}{8 N^4}\right) \nn \\
&&-\f{12}{(4\pi)^4}\left(\frac{363 \gamma_E +394-363 \log 2}{27}-\frac{\Nf (132 \gamma_E +155-132 \log 2)}{27 N} \right. \nn \\
&& \left. +\frac{4\Nf^2 (3 \gamma_E +1-3\log 2)}{27 N^2}+\frac{\Nf}{N^3}\right) \nn \\
& = & \frac{34 N^3-13 N^2 \Nf+3 \Nf}{256 \pi ^4 N^3} = 3\beta_1
\eea
Eq. \eqref{new} reads now:
\bea
\label{C1x2}
g^2_{cd}(\mu)C_1^{(S)}(x) & = & \f{4\beta_0}{\pi^2x^4}g^4_{cd}(\mu)\left(1 + 2\beta_0 g^2_{cd}(\mu)\log(x^2\mu^2) + g^4_{cd}(\mu)(3\beta_1\log(x^2\mu^2) \right. \nn \\
&{}& \left. + 3\beta_0^2 \log^2(x^2\mu^2))\right)
\eea
In order to verify Eq. \eqref{CSS1F2} we insert the identity:
\be
1 = \f{1+\frac{\beta_1}{\beta_0}g^2_{cd}(\mu)}{1+\frac{\beta_1}{\beta_0}g^2_{cd}(x)}\f{1+\frac{\beta_1}{\beta_0}g^2_{cd}(x)}{1+\frac{\beta_1}{\beta_0}g^2_{cd}(\mu)}
\ee
in Eq. \eqref{C1x2} to the relevant perturbative order:
\bea
\label{C1x3}
g^2_{cd}(\mu)C_1^{(S)}(x) & = &  \f{1+\frac{\beta_1}{\beta_0}g^2_{cd}(\mu)}{1+\frac{\beta_1}{\beta_0}g^2_{cd}(x)}\f{1+\frac{\beta_1}{\beta_0}g^2_{cd}(x)}{1+\frac{\beta_1}{\beta_0}g^2_{cd}(\mu)}  \f{4\beta_0 g_{cd}^4(\mu)}{\pi^2 x^4} \nn \\
&& \left(1 + g_{cd}^2(\mu) 2\beta_0 \log(x^2\mu^2) + g_{cd}^4(\mu)( 3\beta_1\log(x^2\mu^2) + 3\beta_0^2 \log^2(x^2\mu^2))\right)  \nn \\
& = & \f{1+\frac{\beta_1}{\beta_0}g^2_{cd}(x)}{1+\frac{\beta_1}{\beta_0}g^2_{cd}(\mu)}\f{4\beta_0 g_{cd}^4(\mu)}{\pi^2 x^4}\left(1 + g_{cd}^2(\mu) 2\beta_0 \log(x^2\mu^2)  \right. \nn \\
&& \left. + g_{cd}^4(\mu)( 2\beta_1\log(x^2\mu^2) + 3\beta_0^2 \log^2(x^2\mu^2))\right) \nonumber \\
& = & \f{4\beta_0}{\pi^2x^4}g^4_{cd}(x)\left(1 + \f{\beta_1}{\beta_0}g^2_{cd}(x) - \f{\beta_1}{\beta_0}g^2_{cd}(\mu)\right)
\eea
where we have expanded to the order of $g^4$:
\be
\f{1+\frac{\beta_1}{\beta_0}g^2_{cd}(\mu)}{1+\frac{\beta_1}{\beta_0}g^2_{cd}(x)} = 1 -  \beta_1 g^4_{cd}(\mu)\log(x^2\mu^2) 
\ee
and to the order of $g^2$:
\be
\f{1+\frac{\beta_1}{\beta_0}g^2_{cd}(x)}{1+\frac{\beta_1}{\beta_0}g^2_{cd}(\mu)} = 1 + \f{\beta_1}{\beta_0}g^2_{cd}(x)-\f{\beta_1}{\beta_0}g^2_{cd}(\mu)
\ee
The factor of $\left(1 + \f{\beta_1}{\beta_0}g^2_{cd}(x) - \f{\beta_1}{\beta_0}g^2_{cd}(\mu)\right)$ in the last line of Eq. \eqref{C1x3} arises according to Eq. \eqref{zeta00}, as it is verified by setting $\gamma^{(F^2)} _{0}=2 \beta_0$ and $\gamma^{(F^2)} _1=4 \beta_1$ in Eq. \eqref{zeta00}:
\bea \label{P1}
&& g^2_{cd}(\mu) C_1^{(S)}(x) \sim   \f{4\beta_0}{\pi^2x^4} g^2_{cd}(x) g^2_{cd}(\mu) Z^{(F^2)}(x \mu, g_{cd}(\mu)) \nonumber \\
&& \sim    \f{4\beta_0}{\pi^2x^4} g^2_{cd}(x) g^2_{cd}(\mu) \left(\frac{g_{cd}(x)}{g _{cd}(\mu)}\right)^{\frac{\gamma_0^{(F^2)}}{\beta_0}} \exp \left( \frac{\gamma_1^{(F^2)}  \beta_0 - \gamma_{0}^{(F^2)} \beta_1}{2 \beta_0^2} (g^2_{cd}(x)-g_{cd}^2(\mu))+\cdots \right) \nonumber \\
&& \sim  \f{4\beta_0}{\pi^2x^4} g^4_{cd}(x) \left(1+ \f{\beta_1}{\beta_0}g^2_{cd}(x) - \f{\beta_1}{\beta_0}g^2_{cd}(\mu) + \cdots \right) 
\eea
Therefore, Eq. \eqref{C1x3} agrees with Eq. \eqref{CSS1F2}. \par
We may remove the scale-dependent term, $- \f{\beta_1}{\beta_0}g^2_{cd}(\mu)$, in the last line of Eqs. \eqref{C1x3} and \eqref{P1} multiplying them by $\beta_0 \left(1 + \f{\beta_1}{\beta_0}g^2_{cd}(\mu)\right) $ that makes $g^2_{cd}(\mu) C_1^{(S)}$ RG invariant to the relevant perturbative order. Hence, we get:
\be
- \f{\beta(g)}{g} C_1^{(S)}(x) \sim \f{4\beta_0^2}{\pi^2 x^4} g^4_{cd}(x)\left(1 + \f{\beta_1}{\beta_0}g^2_{cd}(x)\right) \sim \f{4\beta_0^2}{\pi^2 x^4} g^4_{cd}(x)
\ee
according to the nonperturbative asymptotics in Eq. \eqref{C1UV}.

\end{document}